\documentclass[sigconf, natbib=false, nonacm]{acmart}

\AtBeginDocument{%
  }

\RequirePackage[
  datamodel=acmdatamodel,
  style=acmnumeric,
  ]{biblatex}

\usepackage{graphicx}
\usepackage{amsmath}
\usepackage{multirow}
\usepackage{amsthm}
\usepackage{booktabs}
\usepackage{subcaption}
\usepackage[linesnumbered,ruled,vlined]{algorithm2e}
\usepackage{bm}
\usepackage{enumitem}

\setlist[itemize]{leftmargin = *,topsep=5pt, partopsep=0pt, itemsep = 0pt, parsep=\parskip}

\usepackage{fancyhdr}
\newcommand{\stitle}[1]{\vspace{1ex} \noindent{{\bf #1}}}

\newcommand{\kw}[1]{{\ensuremath {\mathsf{#1}}}\xspace}

\newcommand{\llms}{\kw{LLMs}}
\newcommand{\llm}{\kw{LLM}}
\newcommand{\graphcs}{\kw{GraphCS}}
\newcommand{\csagent}{\kw{CS}-\kw{Agent}}
\newcommand{\solver}{\kw{Solver}}
\newcommand{\validator}{\kw{Validator}}
\newcommand{\decider}{\kw{Decider}}

\newcommand{\up}[1]{\scriptsize(+{#1})}
\begin{document}

\title{CS-Agent: LLM-based Community Search via Dual-agent Collaboration}

\author{Jiahao Hua}
\orcid{0009-0000-9923-8307}
\affiliation{%
  \institution{Nanjing University of Science and Technology}
  \city{Nanjing}
  \country{China}
}
\email{jiahao@njust.edu.cn}

\author{Long Yuan}
\orcid{0000-0001-8111-0401}
\authornote{Corresponding author}
\affiliation{%
  \institution{Wuhan University of Technology}
  \city{Wuhan}
  \country{China}
}
\email{longyuan@whut.edu.cn}

\author{Qingshuai Feng}
\orcid{0009-0001-6985-5352}
\affiliation{%
  \institution{Great Bay University}
  \city{Dongguan}
  \country{China}
}
\email{fengqingshuai@gmail.com}

\author{Qiang Fan}
\orcid{0000-0002-9973-3292}
\affiliation{%
  \institution{National University of Defense Technology}
  \city{Changsha}
  \country{China}
}
\email{fanqiang09@nudt.edu.cn}

\author{Shan Huang}
\orcid{0009-0003-5630-5559}
\affiliation{%
  \institution{The Sixty-third Research Institute, National University of Defense Technology}
  \city{Changsha}
  \country{China}
}
\email{huangshan12@nudt.edu.cn}

\renewcommand{\shortauthors}{Jiahao Hua, et al.}

\begin{abstract}
  Large Language Models (\llms) have demonstrated remarkable capabilities in natural language processing tasks, yet their application to graph structure analysis, particularly in community search, remains underexplored. Community search, a fundamental task in graph analysis, aims to identify groups of nodes with dense interconnections, which is crucial for understanding the macroscopic structure of graphs. In this paper, we propose \graphcs, a comprehensive benchmark designed to evaluate the performance of \llms in community search tasks. Our experiments reveal that while \llms exhibit preliminary potential, they frequently fail to return meaningful results and suffer from output bias. To address these limitations, we introduce \csagent, a dual-agent collaborative framework to enhance \llm-based community search. \csagent leverages the complementary strengths of two \llms acting as \solver and \validator. Through iterative feedback and refinement, \csagent dynamically refines initial results without fine-tuning or additional training. After the multi-round dialogue, \decider module selects the optimal community. Extensive experiments demonstrate that \csagent significantly improves the quality and stability of identified communities compared to baseline methods. To our knowledge, this is the first work to apply \llms to community search, bridging the gap between \llms and graph analysis while providing a robust and adaptive solution for real-world applications.   
\end{abstract}




\maketitle

\section{INTRODUCTION}
Large Language Models (\llms) have achieved remarkable progress in natural language processing (NLP) due to their exceptional capability across diverse tasks such as language generation \cite{kocmi2023findings, lu2023characterised}, question answering \cite{abbasiantaeb2024let, brown2020language, chowdhery2023palm}, and text summarization \cite{ahmed2022few, kolagar2024aligning, tang2023evaluating}. Besides traditional NLP applications, recent studies focus on exploring \llms' capabilities in solving graph-related tasks. For instance, \llms have shown promising results in fundamental graph problems, such as graph reasoning \cite{guo2023gpt4graph, li2024can} and node classification \cite{chen2024exploring, he2023explanations}. This suggests the potential of \llms to handle complex relational data. 
However, their ability to analyze more complex graph structures (i.e., community) remains underexplored.

\begin{figure}[t]
	\centering
	\includegraphics[scale = 0.35]{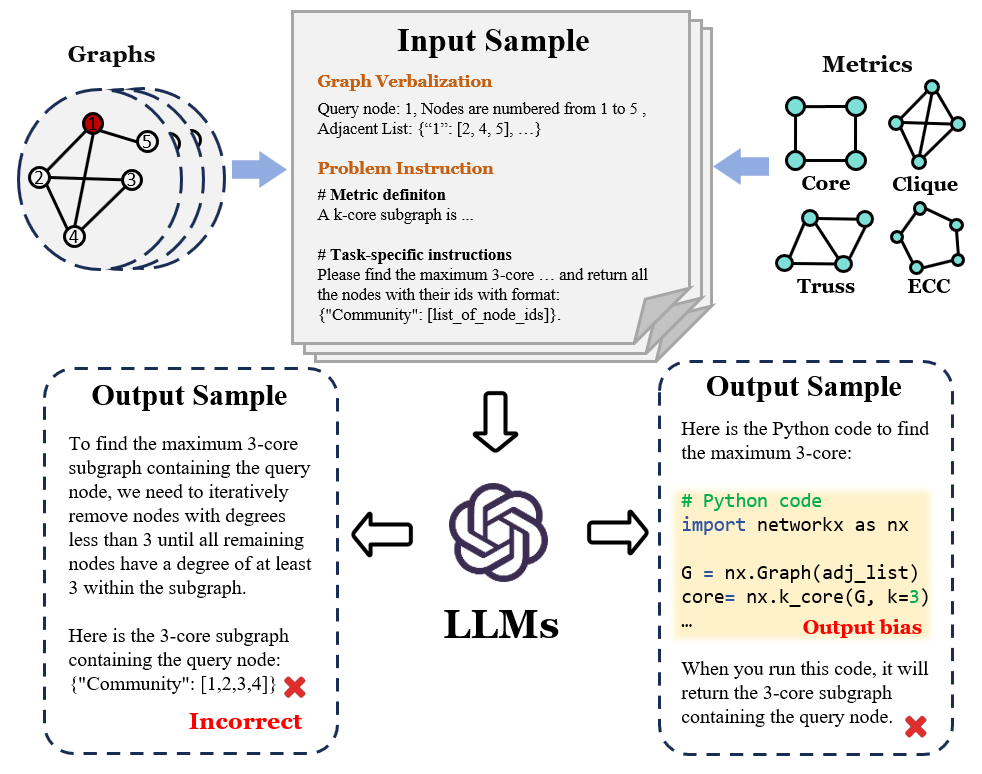}
	\caption{Workflow of LLM-based community search: Given the input graph verbalization and problem instruction, LLMs aim to return a community of the graph that meets specific metric definitions.}
	\label{fig:graphcs}
\end{figure}

Community refers to a cluster of nodes with dense internal connections. These structures often reflect meaningful real-world groupings. For instance, in social networks like Facebook, communities emerge as user groups with shared affiliations or interaction patterns, which represent inherent social organizations \cite{acquisti2006imagined,meng2024survey,yuan2025indexbase}. The task of identifying such communities, known as \textit{Community Search} (\kw{CS}) \cite{huang2017community,DBLP:conf/sigmod/LiuZ0XG20, DBLP:journals/pvldb/LuoLGX25, xu2023effective}, is fundamental in graph analysis. It enables the efficient discovery of groups strongly related to a given node.

In this paper, we aim to explore the latent capability of \llms in discerning hidden community structures within graphs, which presents both challenges and opportunities. A key challenge lies in the scarcity of suitable datasets encompassing diverse CS tasks for comprehensive evaluation. Meanwhile, this gap also enables us to assess whether different \llms exhibit consistent reasoning behaviors. To this end, we introduce \graphcs, a comprehensive benchmark that incorporates two algorithmically generated datasets (PSG and LFR datasets) on four widely-used community search tasks. Additionally, \graphcs employs four prevalent large language models to make systematic comparison of their performance and reasoning patterns in the context of community structure analysis. Our experiments on \graphcs reveal that \llms underperform and generally suffer from \textit{output bias}: their outputs contain specific solutions or codes while missing expected communities. The highlighted part in Figure \ref{fig:graphcs} introduces an example of this bias. These findings emphasize the necessity of a more structured framework to harness the strengths of \llms while addressing their limitations.

Studies in educational psychology \cite{chi2014icap, fawcett2005effect, han2013influence} have demonstrated that collaborative learning through peer interaction can significantly enhance problem-solving efficacy, even when individual participants have limited expertise. Inspired by these findings, we apply similar mechanisms to AI systems and propose \csagent: a dual-agent collaborative framework designed to enhance \llms' capabilities in solving CS tasks. \csagent leverages the strengths of multiple \llms by assigning distinct roles to each model. Specifically, one \llm acts as a \solver, generating candidate communities based on the input graph and query; while another serves as a \validator, evaluating the quality of the generated communities and providing feedback for refinement. The iterative dialogue between the \solver and \validator enables the framework to dynamically improve the quality and stability of the identified communities. Additionally, \csagent comprises a \decider module that selects the optimal community based on a comprehensive evaluation of the candidate results. This module ensures a reliable output of \csagent framework.

To our knowledge, we are the first to apply \llms to community search problem. The \graphcs benchmark and \csagent framework we propose not only bridge the gap between \llms and graph analysis but also provide an adaptable solution for CS tasks across various domains. The main contributions of this paper are as follows: 

\begin{itemize}[leftmargin=*]
    \item We propose a comprehensive benchmark \graphcs to evaluate the potential of \llms in addressing various community search problems. The experimental results reveal that the competence of \llms is limited, with a widespread output bias of failing to return expected results.
    \item We introduce \csagent, a dual-agent collaborative framework where two \llms, acting as \solver and \validator, engage in multi-turn dialogue to dynamically refine community search results. Finally, \decider module selects an optimal community as the output. \csagent provides a robust and flexible approach for community search tasks.
    \item Empirically, \csagent shows remarkable improvements on both quality and stability of the identified communities compared with multiple baselines methods.  
\end{itemize}

\section{PRELIMINARIES}
\stitle{Community search task.} Generally, community search aims to identify high-quality communities in response to query requests. Formally, given a query vertex $q$ in graph $G = (V,E)$, the task involves finding a subgraph $H = (V_H, E_H)$ of $G$ containing $q$ that satisfies two fundamental properties: (1) Connectivity: Ensuring all vertices within the community are interconnected, and (2) Cohesiveness: Maintaining intensive internal connections among vertices according to a specific cohesive metric. 

\stitle{Cohesive metrics.} We introduce four widely-adopted cohesive metrics encompassing typical problem scenarios to evaluate \llms' performance in discovering diverse community structures. The metrics are defined as follows:

\begin{itemize}
	\item \textbf{$k$-Core} \cite{batagelj2003m,liu2020efficient, seidman1983network}: $k$-Core is the largest subgraph of G, in which each vertex's degree is at least $k$ within the subgraph. 
	\item \textbf{$k$-Truss} \cite{chen2021higher, cohen2008trusses,huang2014querying}: $k$-Truss is the largest subgraph of G in which every edge is contained in at least $(k-2)$ triangles within the subgraph.  
	\item \textbf{$k$-Clique} \cite{chen2022balanced, yuan2017index}: $k$-Clique is a set of $k$ vertices of G such that each pair of vertices has an edge. 
	\item \textbf{$k$-Edge Connected Component ($k$-ECC)} \cite{gibbons1985algorithmic,yuan2017efficient}: $k$-ECC is a subgraph of G such that after removing any $k-1$ edges, it is still connected. 
\end{itemize}


\stitle{Categorization of metrics.} The four metrics can be classified into two structural categories: vertex-centric ($k$-Core and $k$-Clique) and edge-centric metrics ($k$-Truss and $k$-ECC). Vertex-centric metrics analyze communities through vertex properties while edge-centric metrics examine edge properties. For instance, $k$-Core quantifies cohesion by minimum vertex degrees while $k$-ECC assesses robustness via edge connectivity. The categorization enables comprehensive evaluation of \llms' CS capabilities across different graph-theoretical perspectives.


\stitle{Problem definition.} LLM-based community search generalizes the traditional CS problem by leveraging \llms to identify community structures. Formally, given graph $G$ and query vertex $q$, \llms generate a connected subgraph $H$ through encoding both the graph topology and predefined cohesiveness metric. The performance of \llms is then evaluated by comparing $H$ against the ground-truth community $H^{*}$.


\section{GRAPHCS BENCHMARK}

To evaluate \llms' performance on CS tasks, datasets comprising sufficient graphs with community structure is necessary. Furthermore, effective task-specific prompts is essential for facilitating \llms' capabilities to solve CS problems. Addressing these fundamental requirements, we propose GraphCS benchmark. This section provides a detailed exposition of its key components.

\begin{table}[htbp]
        \caption{Detailed statistics of two GraphCS benchmark datasets. Column headers describe the total number of graphs in the dataset (Size), range of graph scale ($|V|$), mean edge quantity ($|E|$) and average community size ($|V_H|$).}
	\centering
        \scalebox{0.95}{
	\begin{tabular}{lc|cccc}
		\toprule
		\bfseries Dataset & \bfseries Size  & \bfseries Subset & \bfseries $|V|$ & \bfseries $|E|$  & \bfseries $|V_H|$ \\
		\midrule
		\multirow{3}{*}{PSG Dataset} & \multirow{3}{*}{6,240}
            & $\# Easy$   & 5-10  & 18.6 &  6.1    \\
		& & $\# Medium$ & 11-25  & 56.1  &  8.4 \\
		& & $\# Hard$   & 26-35  & 127.7  & 10.2 \\
		\midrule
            \multirow{3}{*}{LFR Dataset} & \multirow{3}{*}{6,120}
            & $\# Easy$   & 10-20  &  23.2  &  5.5 \\
		& & $\# Medium$ & 21-30  &  75.0  &  9.8 \\
            & & $\# Hard$   & 31-40  &  122.4 & 11.9 \\
		\bottomrule
	\end{tabular}
        }
	\label{tab1}
\end{table}

\subsection{Graph Generate Algorithm}
\label{sec:algor}

\stitle{Read-world datasets fall short on LLM-based CS tasks.} Evaluation of \llms on CS tasks requires datasets with diverse graph structures that simultaneously comply with LLM input constraints. Nevertheless, real-world datasets such as DBLP, Facebook and LiveJournal are characterized by their substantial size \cite{yang2012defining}, far exceeding the input token limits of contemporary \llms. Although subgraph extraction could theoretically mitigate this issue, the inherent complexity and macroscopic community patterns in these networks make it impractical to obtain sufficient well-sized subgraphs for thorough benchmarking. 

Consequently, we employ two synthetic graph generate algorithms to construct tailored datasets with controllable properties. Our datasets include more than 12,400 graphs spanning the four CS tasks. We categorize the graphs into three subsets: \#Easy, \#Medium and \#Hard based on their scale $n$ to control varying algorithmic difficulties. Each subset of different task contains at least 500 graphs, guaranteeing statistically reliable comparisons across tasks. The detailed statistics of two datasets are shown in Table \ref{tab1}.

\stitle{Probabilistic Structural Graph (PSG) Dataset.} To generate a random graph $G = (V,E)$ where V and E denote the set of nodes and edges. Assume that the number of nodes is $n$, thus $V = \{v_1, v_2, ..., v_n\}$, and $|V| = n$. Wang et al. \cite{wang2024can} specify the probability $p$ that there is an edge $e_{ij}$ between two nodes $v_i$ and $v_j$, represented as $P (e_{ij} \in E) = p$. 

To enhance the community structure $H = (V_H, E_H)$ of the generated graphs, we propose an improved algorithm by decomposing the edge probability $p$ into two distinct components: $p_{dense}$ and $p_{sparse}$. Specifically, we randomly select a subset of vertices, denoted as $\mathcal{S}_{dense}$, to represent the core community structure of the graph. The probability of an edge existing between any two vertices within this subset is $p_{dense}$. The remaining pairs of vertices share the probability of $p_{sparse}$. In our experiments, we set $p_{dense} = 0.8$ and $p_{sparse} = 0.2$ for all tasks.

\stitle{LFR Benchmark Datasets.} Considering that the structure of graphs in PSG Dataset are relatively fixed, we employ Lancichinetti-Fortunato-Radicchi (LFR) \cite{lancichinetti2008benchmark} benchmark algorithm to better approximate the topological variability observed in real-world networks. The LFR benchmark is an established method in complex network analysis for generating synthetic graphs with tunable community structures that closely mimic real-world network properties.

In comparison with PSG generator, LFR generator introduces multiple parameters to control the complexity of the generated community. For instance, $tau_1$ and $tau_2$ controls the exponent of the power-law distribution of node degrees and community size respectively, $mu$ determines the proportion of nodes in the network connected to nodes outside the community. We set $tau_1 = 1.8, tau_2 = 1.2, mu = 0.1$ for all graphs. These parameters produce graphs that authentically exhibit hierarchical organization with nested communities. It is particularly suitable for evaluating community search algorithms in realistic scenarios.

\begin{figure*}[ht] 
	\centering
	\includegraphics[scale = 0.45]{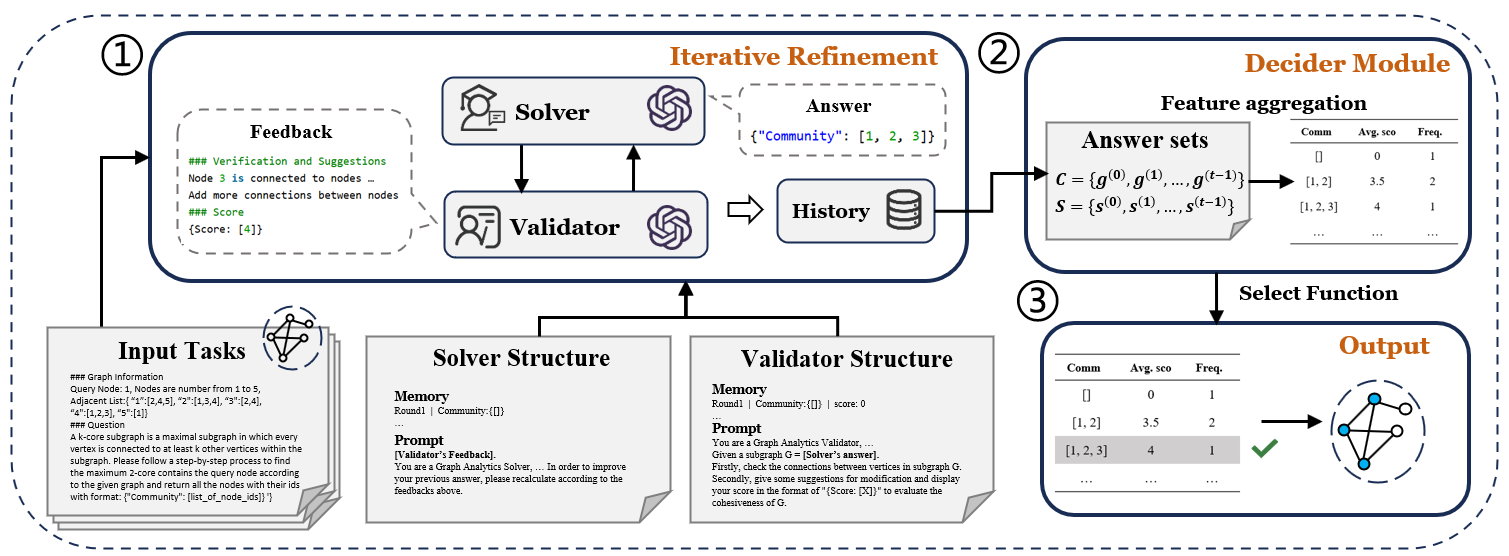}  
	\caption{Framework of CS-Agent.} 
	\label{fig:dac}
\end{figure*}

\subsection{Community Search Prompting}
\label{sec:csprompt}
\subsubsection{Prompting Methods}
To establish comprehensive baselines for evaluating \llms in CS tasks, we adopt a wide range of prompting methods while introducing key adaptations for graph-based problems. These baseline methods include:

\begin{itemize}
    \item \textbf{Zero-shot Prompting (ZERO-SHOT)}: Directly queries the model without task-specific examples, relying on its pre-trained knowledge. 
    \item \textbf{Few-shot in-context learning (FEW-SHOT)} \cite{brown2020language}: Provides a few task examples to guide the model's response generation. In our experiment, Few-shot includes one exemplar randomly selected from our manually constructed samples.
    \item \textbf{Zero-shot chain-of-thought (0-COT)} \cite{kojima2022large}: Encourages step-by-step reasoning without examples to solve complex problems.
\end{itemize}

\subsubsection{Prompt Design}
On the basis of the prompting methods, we develop a standardized task-specific prompt template for \llm-based CS tasks. This structured prompt comprises the following key components:

\stitle{Role Prompting.} Role prompting refers to a technique that assigns a role or persona (e.g., 'scientist' or 'engineer') to an \llm model, thereby influencing both the form and factual precision of their outputs \cite{shanahan2023role}. Prior works have demonstrated that role prompting can significantly improve the performance of \llms \cite{kong2023better, wu2023large}. Building on these findings, we implement specialized role prompts tailored for CS tasks. 

\stitle{Graph Verbalization.} The verbalization of graph topological structure $G_{text}$ through natural language offers an efficient representation paradigm for graph data. Specifically, text-graph $G_{text}$ comprises three key components: (1) Query Node, (2) Node List and (3) Adjacent List. Our representation extends conventional graph encoding methods \cite{jin2024large} by optimizing the format for community search tasks. We verbalize Node List as a sequential enumeration (e.g., \textit{"Nodes are numbered from 0 to n"}), while the Adjacency List adopts a dictionary-style mapping, where each key-value pair denotes a node and its neighboring set, respectively (e.g., \textit{"Adjacent list: \{"0": [1,2]\}"}).

\stitle{Problem Instruction Prompting.} The prompt for problem instruction $p_{inst}$ consists of two elements: (1) Definition of the cohesiveness metric and (2) Task-specific instructions. The definition provides \llms with precise conceptual understanding of the target community structure. While instruction component explicitly directs \llms to perform the required operation while differentiating between baseline approaches.

\subsection{Evaluation Results Analysis}
\label{sec:eva}
We conduct extensive experiments on \graphcs benchmark, with detailed results presented in Section \ref{sec:exp}. Through analysis of the evaluation results, we summarize  the key findings as follows:

\begin{itemize}[leftmargin=*]
    \item \textbf{LLMs possess preliminary community search abilities.} \llms demonstrate impressive performance on some specific tasks, with F1 value exceeding 90\%.
    \item \textbf{Benefits from prompting methods have stark disparity.} 0-CoT has been proved effective for problem-solving. However, its performance in CS tasks is even worse than Zero-shot. In comparison, Few-shot has the best performance among prompting methods.
    \item \textbf{LLMs are brittle to complexity of problems.} Although \llms is competent in easy tasks, their precision dramatically degrades when either the graph size increases or the community structure becomes complicated.
\end{itemize}

Based on the results, an improved framework is necessity to better elicit the community search abilities of \llms.

\section{PROPOSED MODEL}
\subsection{Overview}
This section details our proposed \csagent framework. As illustrated in Figure \ref{fig:dac}, \csagent incorporates two agents, which assume the roles of \solver and \validator respectively. These models engage in multi-round dialogues to evaluate and refine the vertices within the initially identified communities. Upon the conclusion of the dialogue, a \decider module is employed to determine the highest quality community as the final output. Details of the algorithm can be found in Algorithm \ref{alg:dac}.

Each agent is composed of three key components: role prompt, memory, and task instruction. The role prompt specifies agents' role and tasks within the methodology. The memory component retains the historical dialogue content from previous interactions, while task instructions provide agents with explicit guidance on the current task at hand. 

\subsection{Model Details}
\subsubsection{Solver Agent} \solver is proficient in problem-solving, serving as the computational unit within \csagent framework. Its obligation in each conversation round is to output a community conforming to the input requirements. In the initial conversation phase, given the textual representation of the input graph $G_{text}$, problem instruction prompt $p_{inst}$ and \llm model $M$, \solver generates an initial output $y_{sol}^{(0)}$, the included community is denoted as $g^{(0)}$:

\vspace{-1.5mm}
\begin{equation}
	y_{sol}^{(0)} = M(G_{text} || p_{inst})
\label{sol_init}
\end{equation}
\vspace{-1.5mm}

In the iterative refinement process, \solver employs a feedback-driven mechanism to progressively improve its community outputs. In detail, at $t+1$ round of the dialogue, given the memory of the preceding response $mem_{sol}^{(t)}$, feedback $fb^{(t)}$ from \validator regarding the former detected community $g^{(t)}$, along with the optimization prompt $p_{opt}$ and \llm model $M$, \solver generates an enhanced output $y_{sol}^{(t+1)}$, which corresponds to an improved community structure $g^{(t+1)}$ as defined in Equation \ref{sol_iter}. This architectural design empowers \solver to perform targeted improvements addressing specific deficiencies of the communities identified in prior iterations.

\vspace{-2mm}
\begin{equation}
	y_{sol}^{(t+1)} = M(mem_{sol}^{(t)} || fb^{(t)} || p_{upd})
\label{sol_iter}
\end{equation}

\subsubsection{Validator Agent} \validator is dedicated to inspection and proposal, during each round, it appraises the acquired community and comes up with suggestions for modification. Concretely, at iteration t, \validator takes as input: (1) The accumulated evaluation memory $mem_{val}^{(t-1)}$, (2) Graph's textual representation $G_{text}$, (3) Candidate community $g^{(t)}$ proposed by the \solver, (4) the validation prompt template $p_{val}$, and (5) the underlying \llm $M$. The validation process comprises two key phases:

First, \validator conducts a vertex-level examination of $g^{(t)}$, verifying each node's compliance with the topological constraints specified by cohesive metrics (e.g., $k$-Core connectivity requirements). This structural analysis ensures the proposed community satisfies necessary theoretical properties. 

Second, \validator performs a holistic quality assessment, evaluating $g^{(t)}$ against both structural and functional metrics. The output $y_{val}^{(t)}$ encapsulates this dual evaluation through: (1) Structural feedback $fb^{(t)}$: A detailed diagnostic report identifying specific vertices or edges requiring modification, along with concrete suggestions for improvement (e.g., \textit{"Node 7 should be removed as it only has 2 internal connections"}). (2) Quantitative score $score^{(t)} \in [0,5]$: A normalized metric to better quantify and compare the performance of the Solver in identifying high-quality communities. Overall, $y_{val}^{(t)}$ is instantiated as:

\vspace{-1.5mm}
\begin{equation}
	y_{val}^{(t)} = M(mem_{rev}^{(t-1)} || g^{(t)} || p_{val})
\label{valid}
\end{equation} 

\stitle{Cognitive Rigidity Mitigation.} Liang et al. \cite{liang2023encouraging} observe an inherent phenomenon in \llms called Degeneration-of-Thought (DoT), where \llms tend to become increasingly confident in their responses and consequently resistant to self-correction. This cognitive rigidity is particularly problematic for iterative refinement systems, as the retention of the \validator's prior assessments in memory may influence current evaluations. To address this limitation, we implement a strategic optimization for \validator's memory structure. The strategy automatically clear \validator's memory when the following convergence condition is detected: $g^{(t-1)} = g^{(t)}$, where $g^{(t)}$ represents the community structure proposed by the \solver at iteration t. By purging historical evaluation data when solutions stagnate, we prevent cognitive rigidity accumulation and promote \validator to conduct fresh assessments. 

\begin{algorithm}[tb]
	\caption{CS-Agent Framework}
	\label{alg:dac}
	\textbf{Input}: text-graph $G_{text}$; initial prompt $p_{inst}$ \\
	\tcp{Initialization}
	Solver: Initialize prompt for community searching \\
	Validator: Initialize prompt for community reviewing \\
	\While{$t < max\_round$}{
		\tcp{Solver}
            \eIf{t = 0}{
                Generates initial output $y_{sol}^{(0)}$ \\
            }{
                Generates output $y_{sol}^{(t)}$ based on feedback $fb^{(t-1)}$ \\
                Update memory $mem_{sol}^{(t)}$ \\
            }
		\tcp{Validator}
            Validate community $g^{(t)}$ within $y_{sol}^{(t)}$ \\
            Provide feedback $fb^{(t)}$ \\
            Update memory $mem_{val}^{(t)}$ \\
	}
        \tcp{Decider Module}
        Perform feature aggregation on $(\mathcal{C}, \mathcal{S})$ \\
        Decide output $g_{out}$ through select function \\
	\Return{$g_{out}$}
\end{algorithm}

\begin{table*}[htb]
    \caption{Comparative performance evaluation of GraphCS methods on PSG and LFR datasets with multiple LLMs, assessed across four CS tasks categorized by difficulty levels: Easy (E), Medium (M), and Hard (H). The best results are highlighted in bold and the suboptimal results are in underline. }
    \centering
    \fontsize{9}{10}
    \scalebox{0.7}{
        \begin{tabular}{cccccccccccccccccccccccccc}
            \toprule
             \multicolumn{2}{c}{\textbf{Dataset}} & \multicolumn{12}{c}{\textbf{PSG Dataset}} & \multicolumn{12}{c}{\textbf{LFR Dataset}} \\
            \cmidrule(lr){1-2} \cmidrule(lr){3-14} \cmidrule(r){15-26}
        
            \multirow{2}{*}{\textbf{Model}} & \multirow{2}{*}{\textbf{Method}} & \multicolumn{3}{c}{\textbf{$k$-Core}} & \multicolumn{3}{c}{\textbf{$k$-Truss}} & \multicolumn{3}{c}{\textbf{$k$-Clique}} & \multicolumn{3}{c}{\textbf{$k$-ECC}}
            & \multicolumn{3}{c}{\textbf{$k$-Core}} & \multicolumn{3}{c}{\textbf{$k$-Truss}} & \multicolumn{3}{c}{\textbf{$k$-Clique}} & \multicolumn{3}{c}{\textbf{$k$-ECC}} \\
            \cmidrule(lr){3-5} \cmidrule(lr){6-8} \cmidrule(lr){9-11} \cmidrule(lr){12-14}
            \cmidrule(lr){15-17} \cmidrule(lr){18-20} \cmidrule(lr){21-23} \cmidrule(lr){24-26}
            & & E. & M. & H. & E. & M. & H. & E. & M. & H. & E. & M. & H. & E. & M. & H. & E. & M. & H. & E. & M. & H. & E. & M. & H. \\
            \midrule
            \multirow{3}{*}{\textbf{ChatGPT}} 
            & Zero-shot & 82.9 & \underline{66.4} & 52.9 & 73.9 & 49.2 & 33.5 & 66.6 & 53.7 & 34.9 & 67.5 & 36.9 & 15.6 & 64.8 & 56.3 & 30.0 & 61.2 & 40.0 & 18.9 & 62.4 & 51.7 & 24.1 & 33.2 & 24.7 & 18.5 \\
            & Few-shot & 83.8 & \textbf{77.0} & \textbf{79.2} & \underline{83.8} & \textbf{75.7} & 73.2 & 72.4 & \underline{71.3} & \textbf{62.2} & \textbf{83.8} & \textbf{72.3} & \textbf{54.8} & \underline{77.2} & \textbf{69.2} & \textbf{64.1} & \textbf{73.2} & \textbf{62.1} & \textbf{49.9} & 70.6 & \textbf{63.7} & \textbf{48.6} & \underline{53.8} & \textbf{48.7} & \textbf{42.9} \\
            & 0-CoT & 69.2 & 48.8 & 18.1 & 63.6 & 25.5 & 5.7 & 74.1 & \textbf{74.3} & 52.4 & 56.2 & 13.1 & 3.2 & 44.8 & 29.5 & 18.6 & 27.2 & 9.4 & 10.5 & 79.0 & \underline{63.4} & 46.2 & 30.1 & 10.6 & 6.7 \\
            \cmidrule(lr){3-26}
            
            \multirow{3}{*}{\textbf{Gemini}}
            & Zero-shot & 75.5 & 34.6 & 11.5 & 73.4 & 13.3 & 18.7 & 73.6 & 42.3 & 45.5 & 50.0 & 14.7 & 27.9   
            & 55.4 & 27.4 & 8.5 & 31.3 & 16.9 & 4.3 & \underline{80.8} & 59.3 & 43.5 & 9.1 & 15.1 & 25.8 \\
            & Few-shot  & \textbf{93.3} & 36.5 & 19.2 & \textbf{86.6} & 12.5 & 42.9 & \textbf{76.7} & 52.8 & \underline{59.8} & 40.0 & 21.6 & 22.9   
            & \textbf{78.6} & 9.0 & 8.0 & 44.6 & 0.0 & 6.2 & \textbf{87.5} & 59.3 & 43.5 & 10.1 & 17.5 & 5.2 \\
            & 0-CoT     & \underline{90.0} & 12.8 & 8.7 & 19.9 & 1.4 & 3.2 & 40.0 & 14.8 & 11.6 & 23.3 & 2.1 & 0.0 
            & 57.4 & 5.6 & 1.5 & 30.8 & 0.0 & 0.0 & 66.6 & 20.1 & 0.0 & 13.2 & 0.5 & 0.0 \\
            \midrule
            
            \multirow{3}{*}{\textbf{Llama3}}
            & Zero-shot & 65.2 & 43.8 & 27.1 & 51.7 & 21.4 & 12.6 & 32.9 & 37.4 & 19.5 & 61.7 & 46.2 & 19.3 & 44.7 & 33.4 & 24.3 & 39.6 & 15.3 & 17.9 & 42.6 & 22.6 & 26.7 & 33.9 & 25.3 & 22.2 \\
            & Few-shot & 76.2 & 53.7 & \underline{55.0} & 60.6 & 50.0 & 28.4 & 45.6 & 52.8 & 34.9 & \underline{75.2} & \underline{61.8} & 30.1 & 66.4 & 55.0 & 37.2 & 58.7 & 41.7 & 29.6 & 51.2 & 41.4 & 17.8 & 41.4 & 34.1 & 24.2 \\
            & 0-CoT & 55.8 & 36.0 & 31.9 & 49.5 & 37.8 & 6.9 & 55.1 & 26.2 & 20.8 & 63.5 & 33.2 & 6.1 & 53.4 & 35.5 & 30.6 & 27.5 & 22.6 & 7.1 & 50.7 & 30.9 & 21.6 & 28.6 & 19.8 & 4.0 \\
            \cmidrule(lr){3-26}
            
            \multirow{3}{*}{\textbf{Mixtral}}
            & Zero-shot & 79.3 & 48.5 & 32.3 & 77.2 & 49.4 & 36.2 & \underline{76.2} & 52.8 & 44.9 & 70.1 & 58.2 & 35.5  & 75.6 & 47.5 & \underline{44.8} & \underline{68.3} & \underline{41.9} & \underline{42.8} & 74.1 & 52.8 & \underline{47.5} & \textbf{58.3} & \underline{46.1} & 35.8 \\
            & Few-shot  & 78.8 & 59.6 & 35.3 & 79.4 & 55.7 & 41.5 & 65.7 & 53.6 & 33.8 & 70.1 & 43.3 & 29.5  & 48.5 & 24.4 & 28.8 & 46.8 & 19.0 & 16.5 & 75.0 & 61.0 & 36.4 & 33.4 & 44.8 & \underline{39.7} \\
            & 0-CoT     & 77.0 & 48.7 & 28.8 & 81.9 & 42.4 & 24.5 & 67.8 & 51.1 & 35.3 & 66.7 & 52.4 & \underline{39.8}  & 61.4 & 31.1 & 37.1 & 57.6 & 37.3 & 16.7 & 70.8 & 56.3 & 45.7 & 46.1 & 45.6 & 36.1 \\
            \bottomrule
        \end{tabular}}
    \label{graphcs_multi_dataset}
\end{table*}

\subsubsection{Decider Module} After $t$ rounds of iterative refinement, we obtain a set of communities $\mathcal{C} = \{g^{(0)}, g^{(1)}, \dots, g^{(t-1)}\}$ generated by the \solver along with their corresponding scores $\mathcal{S} = \{s^{(0)}, s^{(1)}, \dots, s^{(t-1)}\}$ provided by the \validator. \decider module first performs feature aggregation from $(\mathcal{C}, \mathcal{S})$ to construct a statistical representation of the solution space, denoted by a feature vector $\mathcal{F}$. Then, the optimal community $g_{out}$ is selected via a multi-stage select function.

\stitle{Feature Aggregation.} Before selection, \decider performs feature aggregation on the solution space. First, it merges duplicate communities and for each unique community $g^{(i)}$, it calculates: (1) Average score as the mean of all \validator-assigned scores across different iterations; (2) Occurrence frequency by counting how many times the community appears in $\mathcal{S}$; and (3) Refinement depth recorded as the earliest round index where the community first emerged. These statistics are computed across all topological variants in the solution space, creating a normalized feature profile where each unique community is represented by its aggregated statistical characteristics. 

\stitle{Select Function.} \decider module implements a multi-stage select function that systematically evaluates candidate communities through both primary quantitative metrics and secondary qualitative assessments. The primary selection criterion is based on normalized average scores. This approach prioritizes the \validator's quantitative evaluation as the most reliable indicator of community quality, as the scoring metric incorporates both structural soundness and methodological compliance.

When encountering communities with identical average scores (score ties), \decider implements a hierarchical tie-breaking mechanism based on two secondary criteria: (1) Solution stability: Measured by occurrence frequency, where communities appearing multiple times demonstrate higher algorithmic confidence and convergence reliability; and (2) Refinement depth: Determined by the earliest round of appearance, with later-emerging solutions benefiting from more extensive optimization cycles. Benefiting from selection protocol of the \decider module, the final selection $g_{out}$ represents not only the highest-scoring community but also the most stable and thoroughly optimized solution in the candidate set.

\section{EXPERIMENTAL RESULTS}
\label{sec:exp}

\subsection{Experimental Setup}
\stitle{Language models and hyperparameters.} For backbone language models, we employ ChatGPT (GPT-3.5-turbo) and Gemini (Gemini-2.0-flash) as representative closed-source \llms; Llama3 (Llama3-8B) and Mixtral (Open-mixtral-8x7b) as open-source counterparts, ensuring experimental reproducibility. These models were selected based on their widespread adoption and accessible API interfaces. For all baselines, we set temperature $ \tau = 0.5$; for SC prompting, we set temperature $ \tau = 0.8$ to generate more diverse results.

\stitle{Evaluation Metrics.} F1-score \cite{powers2020evaluation} is a commonly used performance metric which harmonizes both precision and recall. By comparing the predicted community vertices against ground-truth labels, the F1-score provides a unified measure that emphasize the ability of \llms to correctly identify true community members while avoiding the inclusion of irrelevant vertices.

\subsection{LLMs Performance on GraphCS}
We perform comprehensive experiments on the \graphcs benchmark, the results are shown in Table \ref{graphcs_multi_dataset}. The following are some findings we inferred from experimental results.

\subsubsection{ChatGPT outperforms other \llms.} 
Experimental results demonstrate a clear performance hierarchy: closed-source models (ChatGPT, Gemini) outperform open-source counterparts (Llama3, Mixtral). Specifically, ChatGPT consistently achieves the highest accuracy across most tasks when using few-shot prompting, demonstrating robust generalization capabilities. As for Gemini, while it shows excellent performance on easy-level tasks (e.g., 93.3\%/$k$-Core and 86.6\%/$k$-Truss in PSG with Few-shot) but suffers significant scalability limitations (declining to 0\% on several hard-level tasks). In contrast, open-source models exhibit lower sensitivity to prompting strategies. Mixtral in particular shows minimal performance variation across different prompting methods, suggesting limited adaptability to task-specific cues.

Considering cost-efficiency and practical deployment constraints, we need to select a most representative \llm for subsequent experiments on \csagent. ChatGPT's versatility and stability has been proven in complex reasoning tasks \cite{chang2024survey}, along with its superior performance on the \graphcs benchmark with its strong adaptability to prompting methods and clear differentiation in task difficulty. To this end, we select GPT-3.5-turbo as the base \llm for further experimentation.

\subsubsection{Few-shot demonstrates superior performance.}
Few-shot prompting method presents more stable performance across different tasks and \llms models, outperforming both Zero-shot and 0-CoT methods. Few-shot method exhibits strong robustness across graphs of varying scales, indicating that in-context examples effectively anchor \llms' behavior by providing both structural guidance for output formatting and task-specific reasoning patterns. Conversely, Zero-shot and 0-CoT methods pronounce substantially greater sensitivity to graph complexity, with 0-CoT proving particularly ineffective - achieving F1 scores below 10\% on most hard tasks. This suggests that the conventional chain-of-thought reasoning approach, while effective in other domains, fails to provide meaningful improvements for CS tasks.

\subsubsection{Vertex-centric tasks maybe easier for \llms.}
As previously discussed, we have classified the four CS metrics into two distinct categories: vertex-centric ($k$-Core, $k$-Clique) and edge-centric ($k$-Truss, $k$-ECC) tasks. Our experimental results reveal that \llms address vertex-centric tasks better than edge-centric tasks. Interestingly, this observed phenomenon contrasts with the computational complexity of conventional graph algorithms. Detection of $k$-Clique subgraphs has the highest time complexity among four tasks \cite{fang2020survey}, while \llms demonstrate relatively robust performance on this task. This paradox suggests that \llms likely employ fundamentally different reasoning mechanisms compared to traditional graph algorithms. The superior performance of \llms on vertex-centric tasks indicates their enhanced capability in identifying and leveraging vertex-associated topological features, which may stem from \llms' inherent strength in pattern recognition and relational learning.

\subsubsection{Graph characteristics have significant effect.} \graphcs incorporates two key graph characteristics: (1) Graph scale and (2) Community structure topology. The results demonstrate an inverse correlation between graph scale and method accuracy, with performance degradation becoming particularly pronounced in Zero-shot and 0-CoT approaches. This effect suggests fundamental limitations in \llms' capacity to process large-scale topological data without explicit structural guidance. Furthermore, PSG and LFR datasets comprise different community topology, with graphs in PSG datasets exhibiting uniform degree distributions, while LFR benchmark introducing realistic network properties. Results reveal that methods generally have limited performance on LFR dataset, demonstrating that community topology also has a certain impact on the performance of the methods.

\begin{figure}[htbp]
    \centering
    \captionsetup[subfigure]{font=scriptsize}
    \begin{subfigure}[b]{0.2\textwidth}
        \centering
        \includegraphics[width=\linewidth]{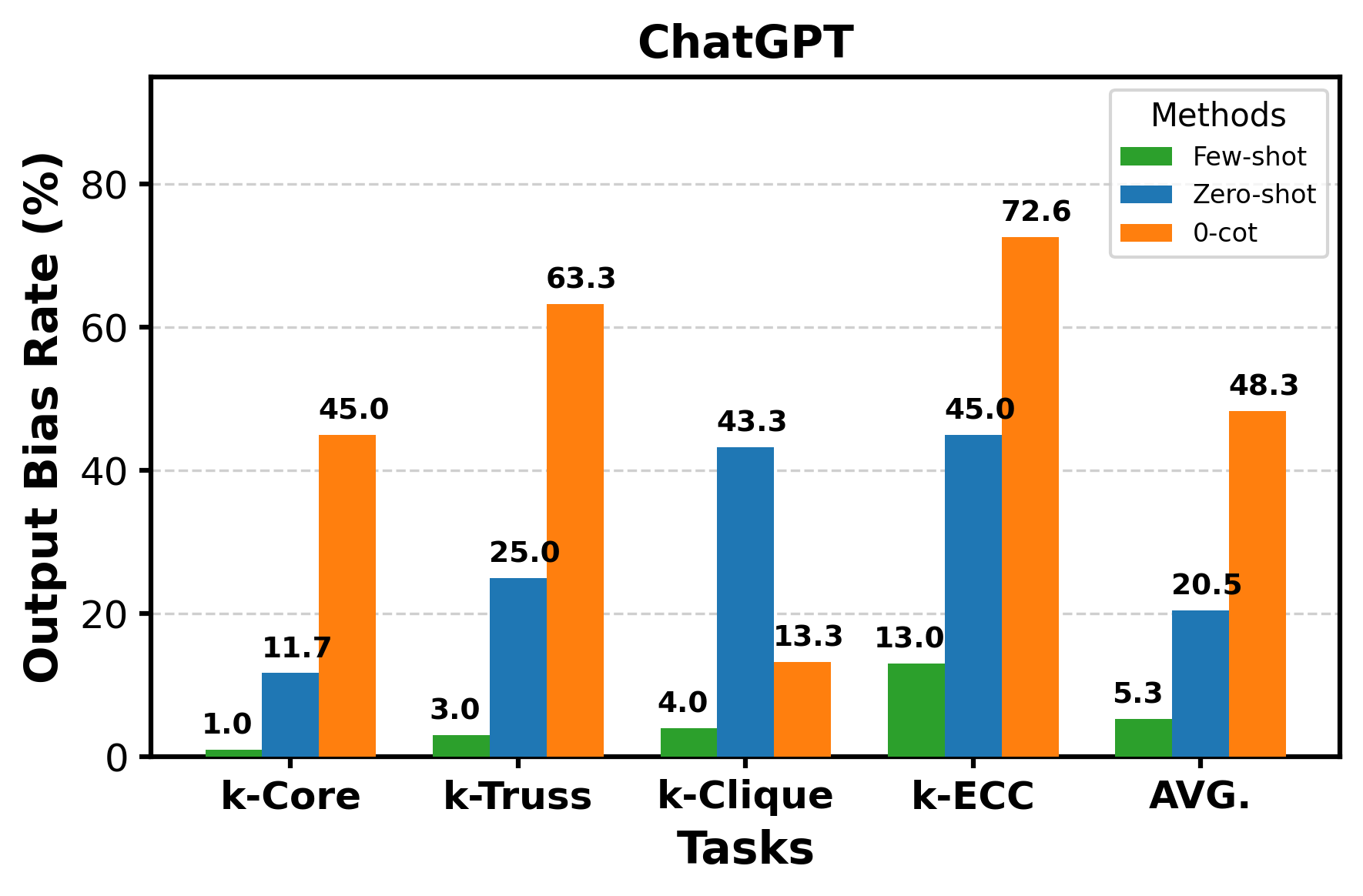}
        \caption{ChatGPT on PSG Dataset}
    \end{subfigure}
    \hspace{0.005\textwidth} 
    \begin{subfigure}[b]{0.2\textwidth}
        \centering
        \includegraphics[width=\linewidth]{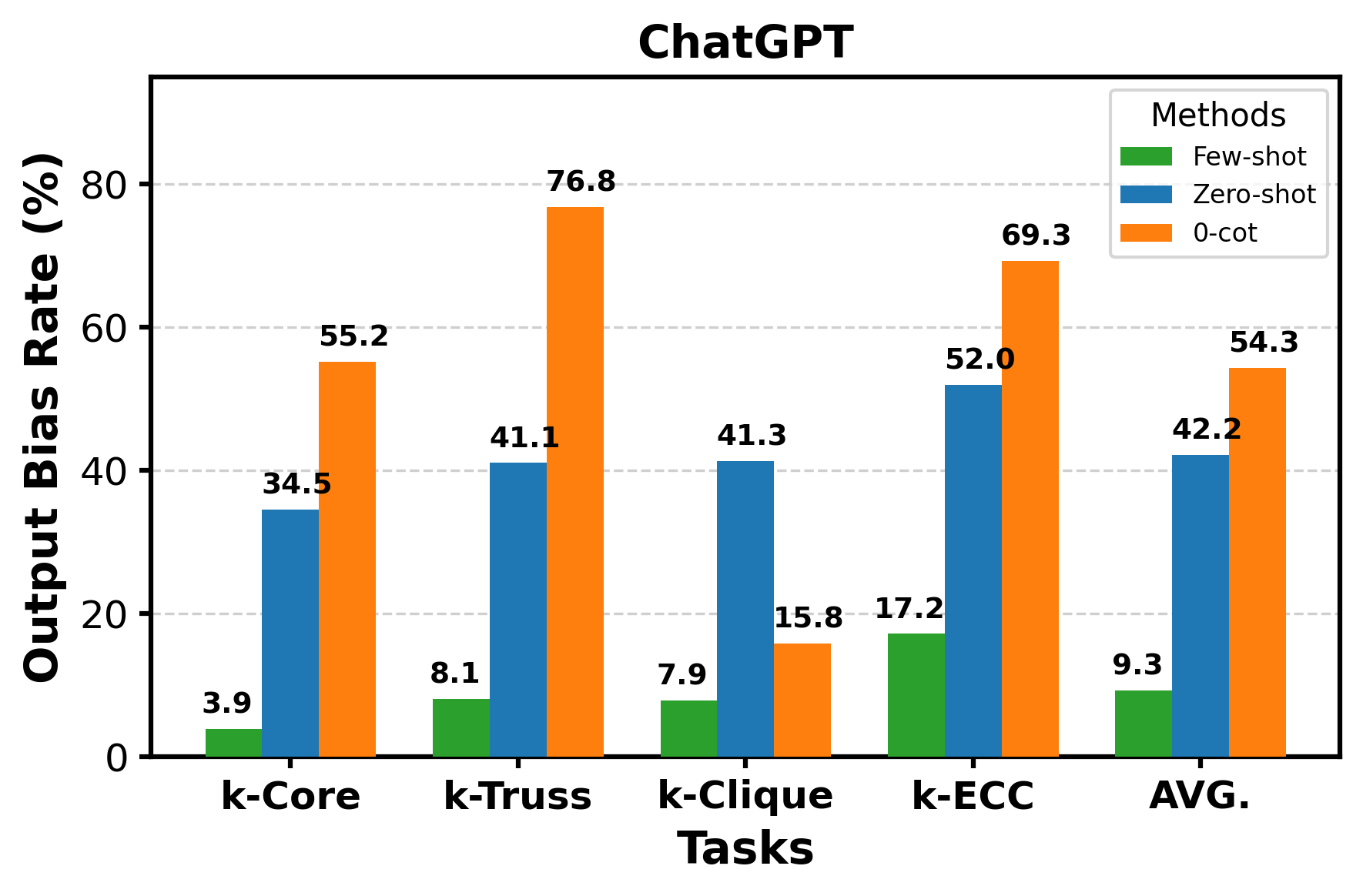} 
        \caption{ChatGPT on LFR Dataset}
    \end{subfigure}
    
    \begin{subfigure}[b]{0.2\textwidth}
        \centering
        \includegraphics[width=\linewidth]{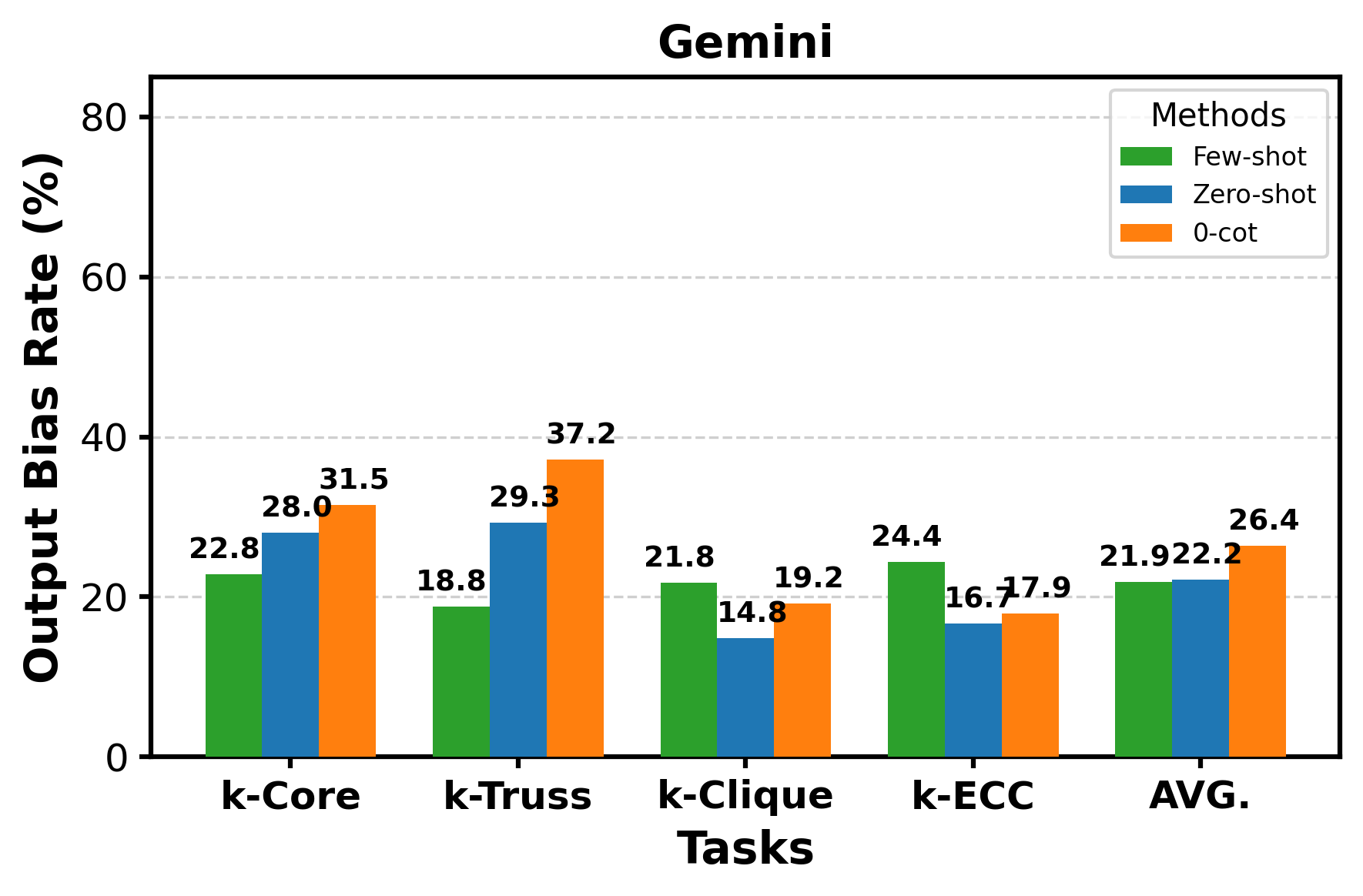}
        \caption{Gemini on PSG Dataset}
    \end{subfigure}
    \hspace{0.005\textwidth} 
    \begin{subfigure}[b]{0.2\textwidth}
        \centering
        \includegraphics[width=\linewidth]{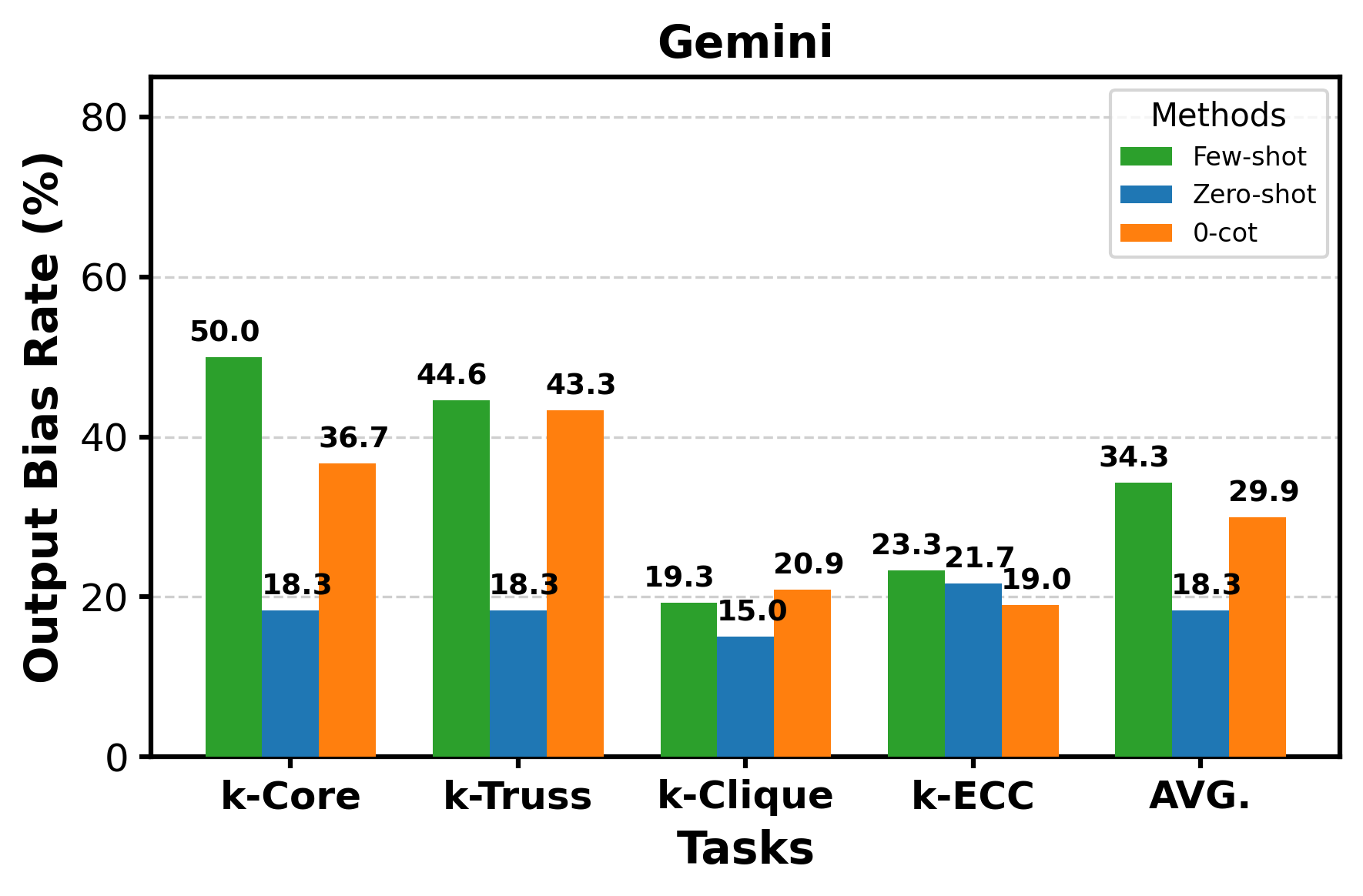}
        \caption{Gemini on LFR Dataset}
    \end{subfigure}

    \begin{subfigure}[b]{0.2\textwidth}
        \centering
        \includegraphics[width=\linewidth]{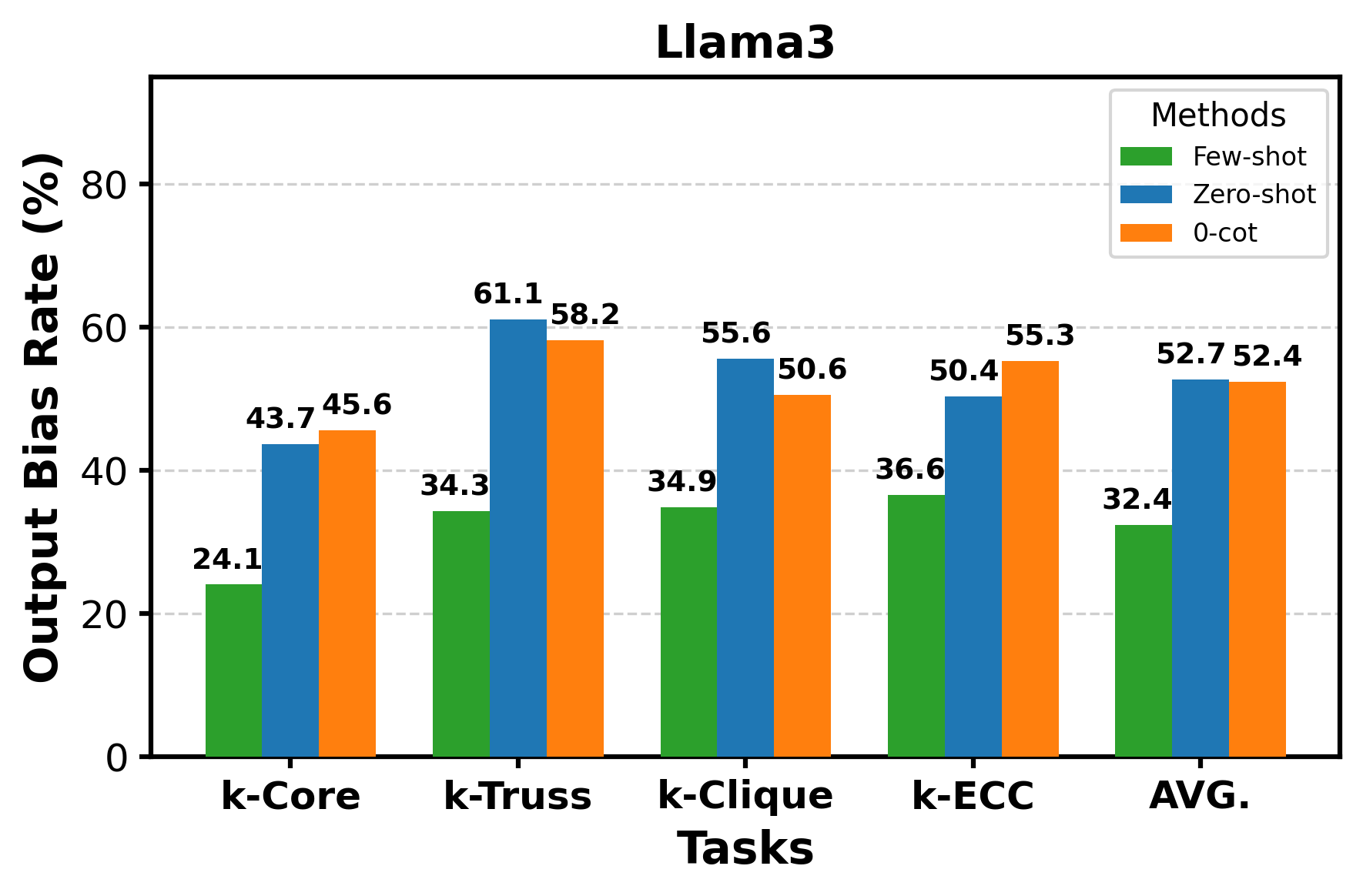}
        \caption{Llama3 on PSG Dataset}
    \end{subfigure}
    \hspace{0.005\textwidth} 
    \begin{subfigure}[b]{0.2\textwidth}
        \centering
        \includegraphics[width=\linewidth]{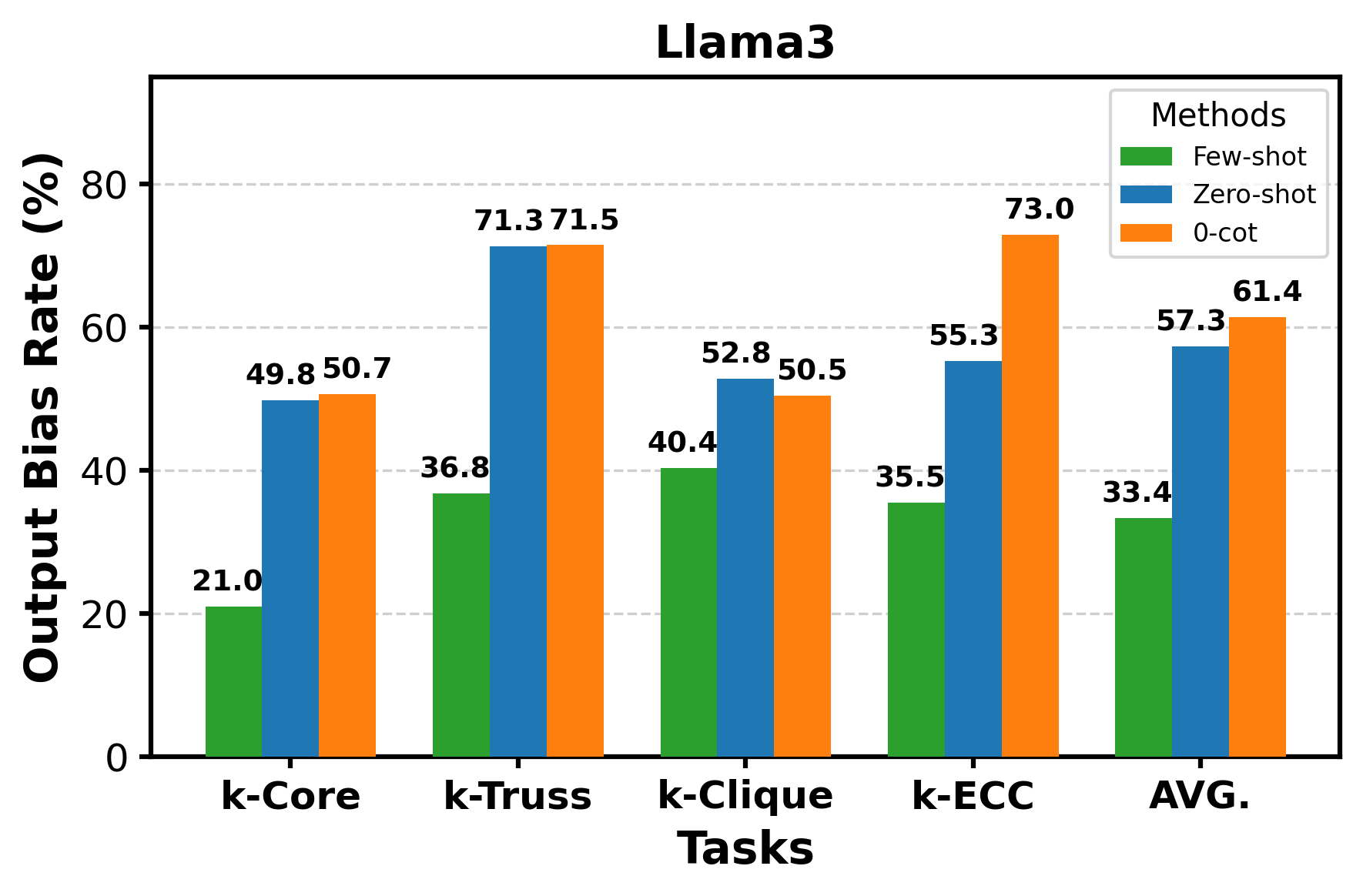}
        \caption{Llama3 on LFR Dataset}
    \end{subfigure}

    \begin{subfigure}[b]{0.2\textwidth}
        \centering
        \includegraphics[width=\linewidth]{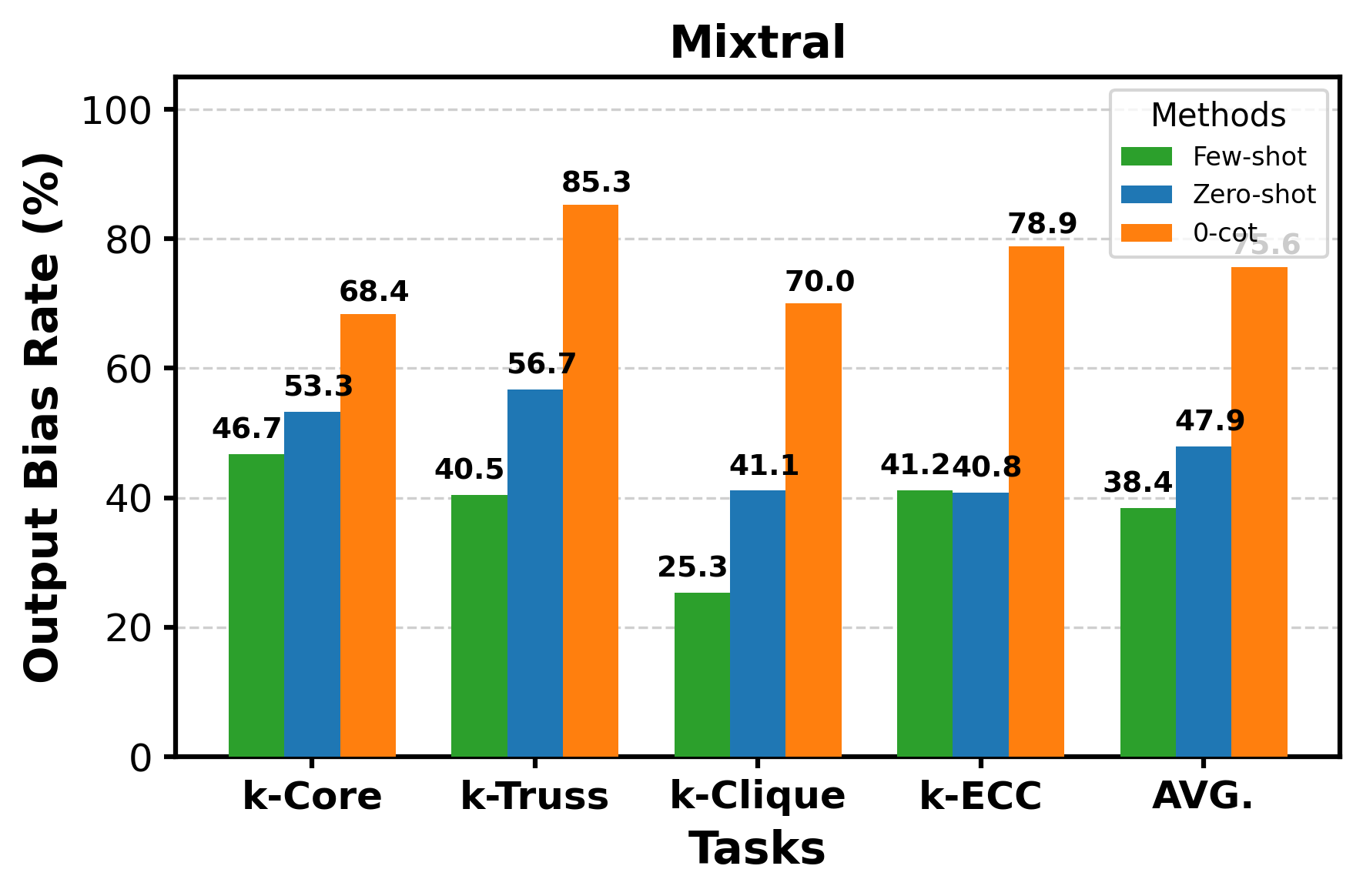}
        \caption{Mixtral on PSG Dataset}
    \end{subfigure}
    \hspace{0.005\textwidth} 
    \begin{subfigure}[b]{0.2\textwidth}
        \centering
        \includegraphics[width=\linewidth]{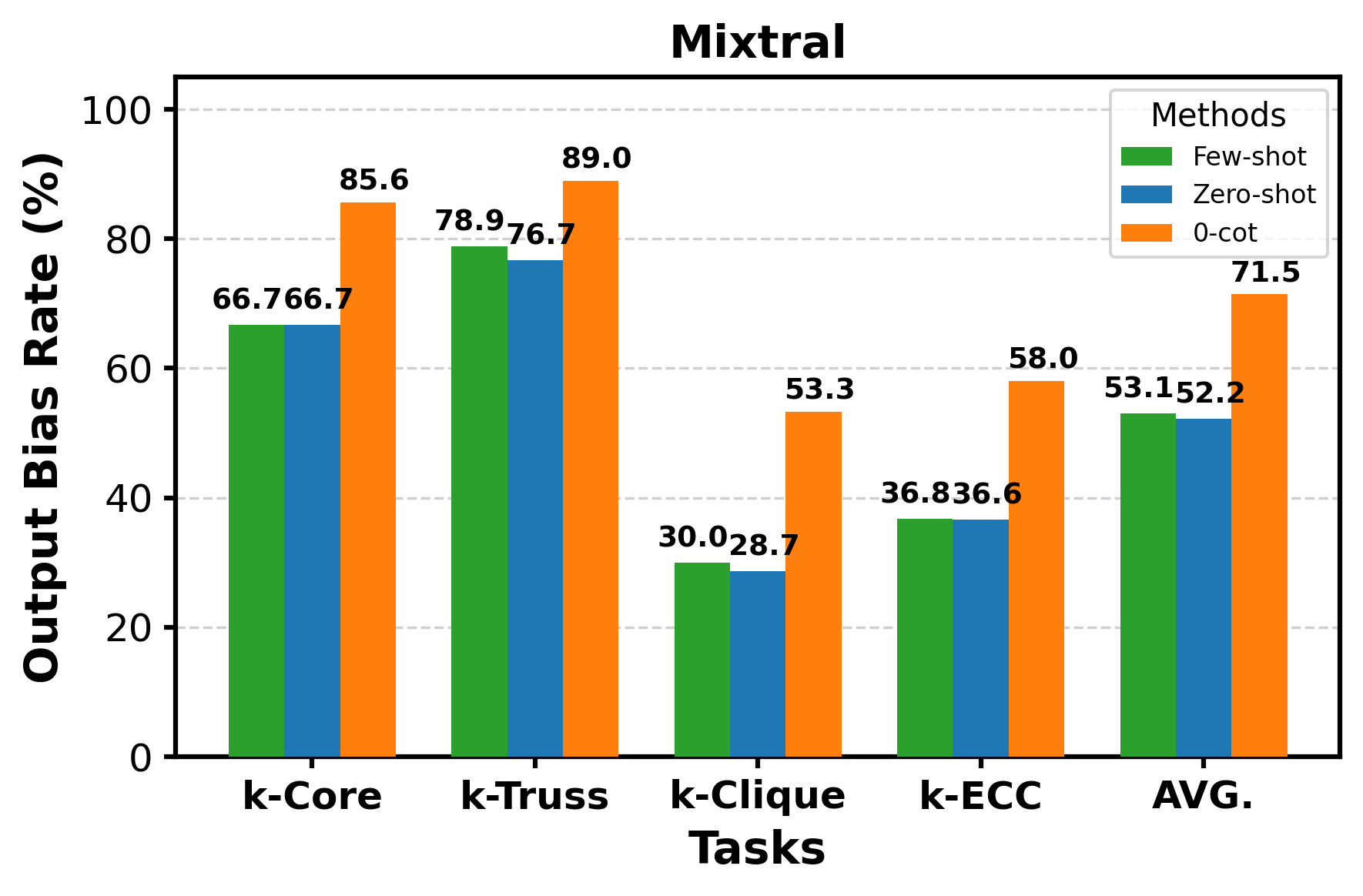}
        \caption{Mixtral on LFR Dataset}
    \end{subfigure}  

    \caption{Output bias proportion over distinct LLMs models and tasks.} 
    \label{fig:POB}
\end{figure}

\begin{table*}[htb]
    \caption{Performance improvement of the CS-Agent over GraphCS benchmark methods using ChatGPT, where $\Delta$ F1 (+) quantifies the accuracy enhancement attributable to CS-Agent. We set dialogue rounds parameter r=3.}
	\centering
        \fontsize{9}{10}
	\scalebox{0.78}{
		\begin{tabular}{ccllllllllllll}
			\toprule
			\multirow{2}{*}{\textbf{Dataset}} & \multirow{2}{*}{\textbf{Method}} & \multicolumn{3}{c}{\textbf{$k$-Core}} & \multicolumn{3}{c}{\textbf{$k$-Truss}} & \multicolumn{3}{c}{\textbf{$k$-Clique}} & \multicolumn{3}{c}{\textbf{$k$-ECC}} \\
            \cmidrule(lr){3-5} \cmidrule(lr){6-8} \cmidrule(lr){9-11} \cmidrule(lr){12-14}
             & & E. & M. & H. & E. & M. & H. & E. & M. & H. & E. & M. & H. \\
		      \midrule
		    \multirow{6}{*}{\textbf{\shortstack{PSG\\Dataset}}} & Zero-shot & 82.9 & 66.4 & 52.9 & 73.9 & 49.2 & 33.5 & 66.6 & 53.7 & 34.9 & 67.5 & 36.9 & 15.6  \\
                 & \textbf{+CS-Agent} & \textbf{87.3} \up{4.4} & \textbf{80.1} \up{13.7} & \textbf{83.0} \up{30.1} & \textbf{85.2} \up{11.3} & \textbf{78.6} \up{29.4} & \textbf{75.2} \up{41.7} & \textbf{81.2} \up{14.6} & \textbf{81.8} \up{28.1} & \textbf{73.9} \up{39.0} & \textbf{86.3} \up{18.8} & \textbf{78.1} \up{41.2} & \textbf{77.2} \up{61.6} \\
                \cmidrule(lr){2-14}
                & Few-shot      & 83.8 & 77.0 & 79.2 & 83.8 & 75.7 & 73.2 & 72.4 & 71.3 & 62.2 & 83.8 & 72.3 & 54.8 \\
                & \textbf{+CS-Agent} & \textbf{85.5} \up{1.7} & \textbf{78.2} \up{1.2} & \textbf{79.3} \up{0.1} & \textbf{86.3} \up{2.5} & \textbf{78.7} \up{2.0} & \textbf{75.6} \up{2.4} & \textbf{74.0} \up{1.6} & \textbf{72.3} \up{1.0} & \textbf{64.6} \up{2.4} & \textbf{86.4} \up{2.6} & \textbf{77.2} \up{5.1} & \textbf{77.5} \up{22.7} \\
                \cmidrule(lr){2-14}
                & 0-CoT         & 69.2 & 48.8 & 18.1 & 63.6 & 25.5 & 5.7 & 74.1 & 74.3 & 52.4 & 56.2 & 13.1 & 3.2 \\
                & \textbf{+CS-Agent} & \textbf{85.0} \up{15.8} & \textbf{78.2} \up{29.4} & \textbf{78.2} \up{60.1} & \textbf{86.2} \up{22.6} & \textbf{79.7} \up{54.2} & \textbf{71.3} \up{65.6} & \textbf{77.2} \up{3.1} & \textbf{79.1} \up{4.8} & \textbf{72.9} \up{20.5} & \textbf{88.1} \up{31.9} & \textbf{76.9} \up{63.8} & \textbf{76.7} \up{73.5} \\
                \midrule
                 \multirow{6}{*}{\textbf{\shortstack{LFR\\ Dataset}}} & Zero-shot & 64.8 & 56.3 & 30.0 & 61.2 & 40.0 & 18.9 & 62.4 & 51.7 & 24.1 & 33.2 & 24.7 & 18.5 \\
                 & \textbf{+CS-Agent} & \textbf{80.9} \up{16.1} & \textbf{71.6} \up{15.3} & \textbf{68.3} \up{38.3} & \textbf{74.8} \up{13.6} & \textbf{65.2} \up{25.2} & \textbf{61.6} \up{42.7} & \textbf{75.6} \up{13.2} & \textbf{72.7} \up{21.0} & \textbf{58.3} \up{24.2} & \textbf{62.4} \up{29.2} & \textbf{60.3} \up{35.6} & \textbf{54.2} \up{35.7} \\
                \cmidrule(lr){2-14}
                & Few-shot   & 77.2 & 69.2 & 64.1 & 73.2 & 62.1 & 49.9 & 70.6 & 63.7 & 48.6 & 53.8 & 48.7 & 42.9 \\
                & \textbf{+CS-Agent} & \textbf{78.9} \up{1.7} & \textbf{71.1} \up{1.9} & \textbf{68.8} \up{4.7} & \textbf{75.9} \up{3.7} & \textbf{65.7} \up{2.6} & \textbf{58.9} \up{9.0} & \textbf{73.1} \up{2.5} & \textbf{65.2} \up{2.5} & \textbf{50.7} \up{2.1} & \textbf{63.2} \up{9.4} & \textbf{57.0} \up{8.3} & \textbf{56.6} \up{13.7} \\
                \cmidrule(lr){2-14}
                & 0-CoT         & 44.8 & 29.5 & 18.6 & 27.2 & 9.4 & 10.5 & 79.0 & 63.4 & 46.2 & 30.1 & 10.6 & 6.7 \\
                & \textbf{+CS-Agent} & \textbf{77.2} \up{32.4} & \textbf{68.5} \up{39.0} & \textbf{64.8} \up{46.2} & \textbf{76.3} \up{49.1} & \textbf{66.5} \up{57.1} & \textbf{63.6} \up{53.1} & \textbf{79.4} \up{0.4} & \textbf{70.2} \up{6.8} & \textbf{63.6} \up{17.4} & \textbf{62.6} \up{32.5} & \textbf{60.8} \up{50.2} & \textbf{53.9} \up{47.2} \\
			\bottomrule
	   \end{tabular}}
	\label{tab3}
\end{table*}

\subsection{Output Bias} 
Previous analysis reveals Zero-shot and 0-CoT methods exhibit significant performance limitations. To further investigate the underlying reasons, particularly why step-by-step reasoning does not work on CS tasks, we conduct a systematic examination of \llms outputs. We identify a critical behavioral pattern: when employing Zero-shot and 0-CoT prompting, \llms predominantly generate specific solutions or codes to solve the problem instead of returning the expected communities. We formally define this pattern as \textit{Output Bias}, directly contributes to the observed accuracy degradation.

\subsubsection{Output bias is universal over \llms models.} 
Figure \ref{fig:POB} quantifies output bias across tested \llms, revealing this as a fundamental challenge in LLM-based CS tasks rather than being model-specific. Statistically, Gemini shows the lowest bias rates, followed by ChatGPT, while Llama3 and Mixtral have higher average rates exceeding 30\%. Except architectural differences in robustness, results further highlight significant variations across prompting methods. In general, 0-CoT produces the highest ratio, while Few-shot has the lowest. This is most prominently manifested in ChatGPT, which displays the greatest variability across prompting methods: 0-CoT yields high bias rates (48.3\% and 54.3\%), whereas Few-shot performs significantly better (5.3\% and 9.3\%). These observations support our hypothesis that the step-by-step reasoning approach in 0-CoT tends to over-emphasize intermediate solution steps or code segments at the cost of final output correctness, thereby exacerbating output bias. Although Few-shot demonstrates better performance through example-based guidance, it still remains susceptible to output bias. Collectively, the following results underscore the need for architectural innovations specifically addressing positional encoding robustness in \llm-based CS tasks.

\subsection{Enhancement of CS-Agent}
Through iterative feedback and refinement, \csagent aims to alleviate the output bias existing in various methods, thereby optimizing the performance on solving \llm-based CS problems. We conducted experiments based on \graphcs benchmark to evaluate the enhancement of \csagent.

\subsubsection{Performance Overview}
Through experiments, we found that \csagent effectively eliminates output bias across methods and greatly improves performance. As presented in Table \ref{tab3}, all baseline prompting methods exhibit enhanced performance when adapted with \csagent. Specifically, 0-CoT and Zero-shot achieve the most substantial gains of 73.5\% and 61.6\% in the hard $k$-ECC task of PSG respectively. Similar improvements on the LFR dataset confirm the framework's generalization across graph structures. It implies that \csagent's multi-round dialogue and refinement process effectively enables \llms to identify and rectify inherent output bias while preserving each method's distinctive prompting characteristics.

Furthermore, Few-shot also benefits from \csagent, albeit to a lesser extent due to its inherently lower bias rates (1.0\% and 3.9\% for $k$-Core tasks). They still achieve measurable gains: improved by 1.7\% in the easy $k$-Core task of PSG Dataset and 4.7\% in the hard $k$-Core of LFR Dataset. This indicates that \csagent can not only eliminate bias but also stimulate the potential of \llms for self-correction. These results underscore the efficacy of \csagent in enhancing the performance of baseline models and highlight the potential of \llms to refine their outputs through structured interaction and feedback.

\begin{table}
    \caption{Performance comparison of Zero-shot based CS-Agent and self-consistency method on GraphCS benchmark. We set candidate number k = 3.}
    \centering
    \fontsize{9.5}{10}
    \scalebox{0.9}{
        \begin{tabular}{llcccc}
            \toprule
            \multicolumn{2}{c}{\textbf{Dataset}} & \multicolumn{2}{c}{\textbf{PSG Dataset}} & \multicolumn{2}{c}{\textbf{LFR Dataset}} \\
            \cmidrule(lr){1-2} \cmidrule(lr){3-4} \cmidrule(lr){5-6}
            \textbf{Metrics} & \textbf{Subset} & \textbf{SC} & \textbf{CS-Agent} & \textbf{SC} & \textbf{CS-Agent} \\
            \midrule
            \multirow{3}{*}{$k$-Core} & \#Easy & 83.2 & \textbf{87.3}  & 76.1 & \textbf{80.9}  \\
            & \#Medium & 63.4 & \textbf{80.1} & 49.0 & \textbf{71.6}  \\
            & \#Hard & 50.2 & \textbf{83.0}  & 34.7 & \textbf{68.3} \\
            \midrule
            \multirow{3}{*}{$k$-Truss} & \#Easy & 80.7 & \textbf{85.2}  & 55.9 & \textbf{74.8}  \\
            & \#Medium & 59.4 & \textbf{78.6}  & 48.1 & \textbf{65.2}  \\
            & \#Hard & 35.0 & \textbf{75.2}  & 27.9 & \textbf{61.6}  \\
            \midrule
            \multirow{3}{*}{$k$-Clique} & \#Easy & 62.9 & \textbf{81.2}  & 71.9 & \textbf{75.6}  \\
            & \#Medium & 52.4 & \textbf{81.8}  & 47.6 & \textbf{72.7}  \\
            & \#Hard   & 31.8 & \textbf{73.9}  & 29.5 & \textbf{58.3}  \\
            \midrule
            \multirow{3}{*}{$k$-ECC} & \#Easy  & 73.9 & \textbf{86.3} & 28.0 & \textbf{62.4} \\
            & \#Medium & 25.4 & \textbf{78.1}  & 15.9 & \textbf{60.3}  \\
            & \#Hard & 9.8 & \textbf{77.2}   & 17.4 & \textbf{54.2} \\
            \bottomrule
        \end{tabular}}
    \label{tab:sc}
\end{table}

\subsection{Effectiveness Analysis}
In this section, in order to investigate the operational efficacy of \csagent, we conducted a comprehensive ablation study targeting its core components.

\subsubsection{Effectiveness of \textit{Validator}} 
The role of the \validator in \csagent framework is to evaluate the quality of the community obtained by \solver and provide feedback. In order to verify the effectiveness of the feedback, we analyze the bias between the F1 value of the communities obtained in each round of dialogue and the review score obtained from \validator's feedback, as illustrated in Figure \ref{fig:bias}. To clearly track bias evolution trend, we choose Zero-shot as the prompting method for the experiment. As evidenced by the figure, the curves of F1 value and review score almost exhibit consistent trends in four tasks, reflecting minimal bias. 

For instance, when solving $k$-Core tasks, Zero-shot perform well in the first round, which corresponds to a relatively high review score close to 3.0. Afterwards, F1 value of the communities gradually decrease after the second round, the review score curve also shows a downward trend at this time. Conversely, Zero-shot's performance on $k$-ECC tasks is notably poorer, with an F1 value just over 50\%, and the \validator assigned a significantly lower score below 2.0. In the second round, both the F1 value and the review score show a significant increase. This progression underscores the \validator's ability to effectively evaluate the results generated by the \solver and provide constructive feedback, thereby guiding the \solver towards more accurate and higher-quality community search progress. These findings substantiate the critical role of the \validator in the iterative refinement process, ensuring continuous improvement in the quality of the detected communities.

\subsubsection{Effectiveness of Decider module} 
\begin{figure}[h]
    \centering
    \captionsetup[subfigure]{font=scriptsize}
    
    \begin{subfigure}[b]{0.23\textwidth}
        \centering
        \includegraphics[width=\linewidth]{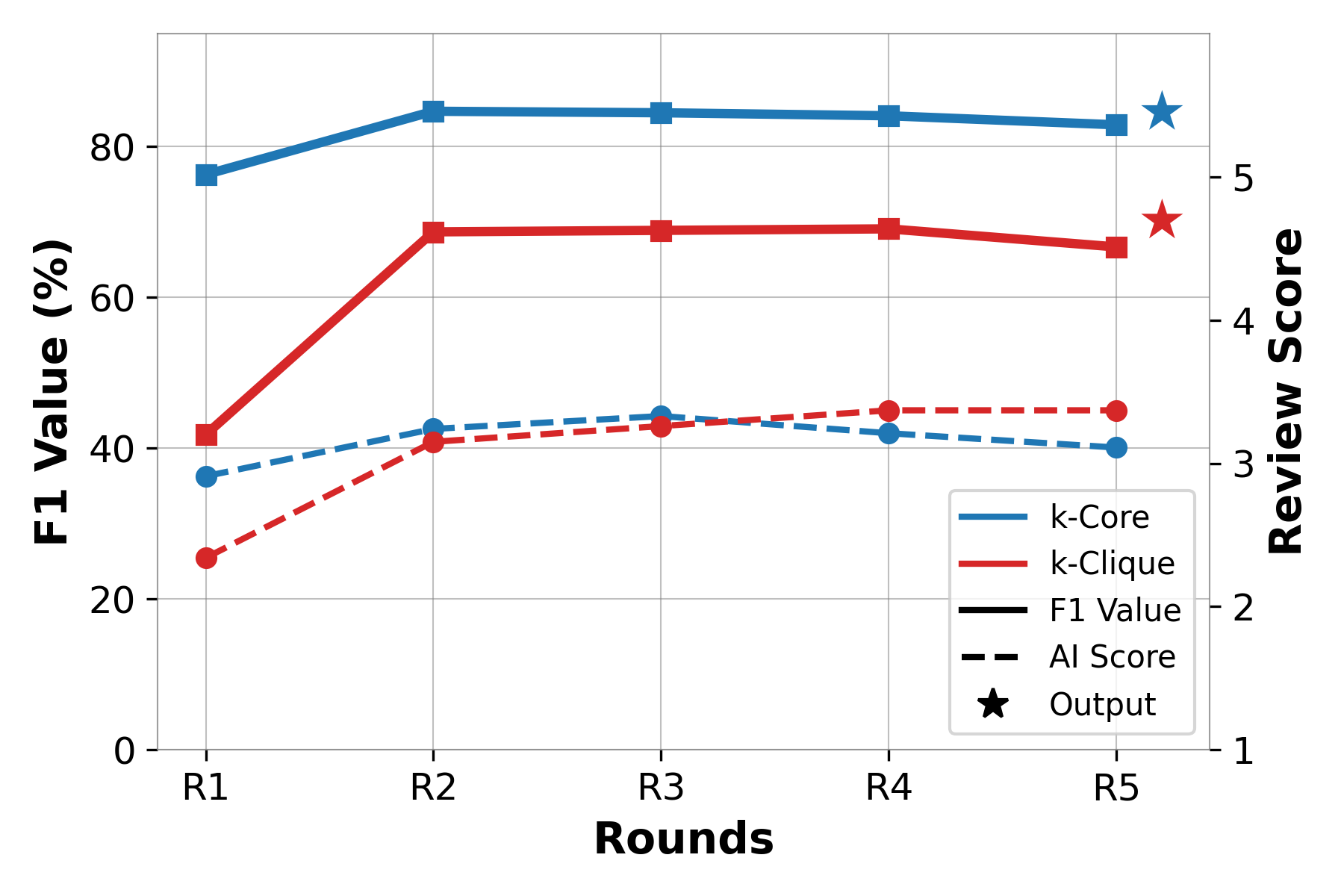}
        \caption{$k$-Core and $k$-Clique}
        \label{fig:zero_core_clique}
    \end{subfigure}
    \hspace{0\textwidth}
    \begin{subfigure}[b]{0.23\textwidth}
        \centering
        \includegraphics[width=\linewidth]{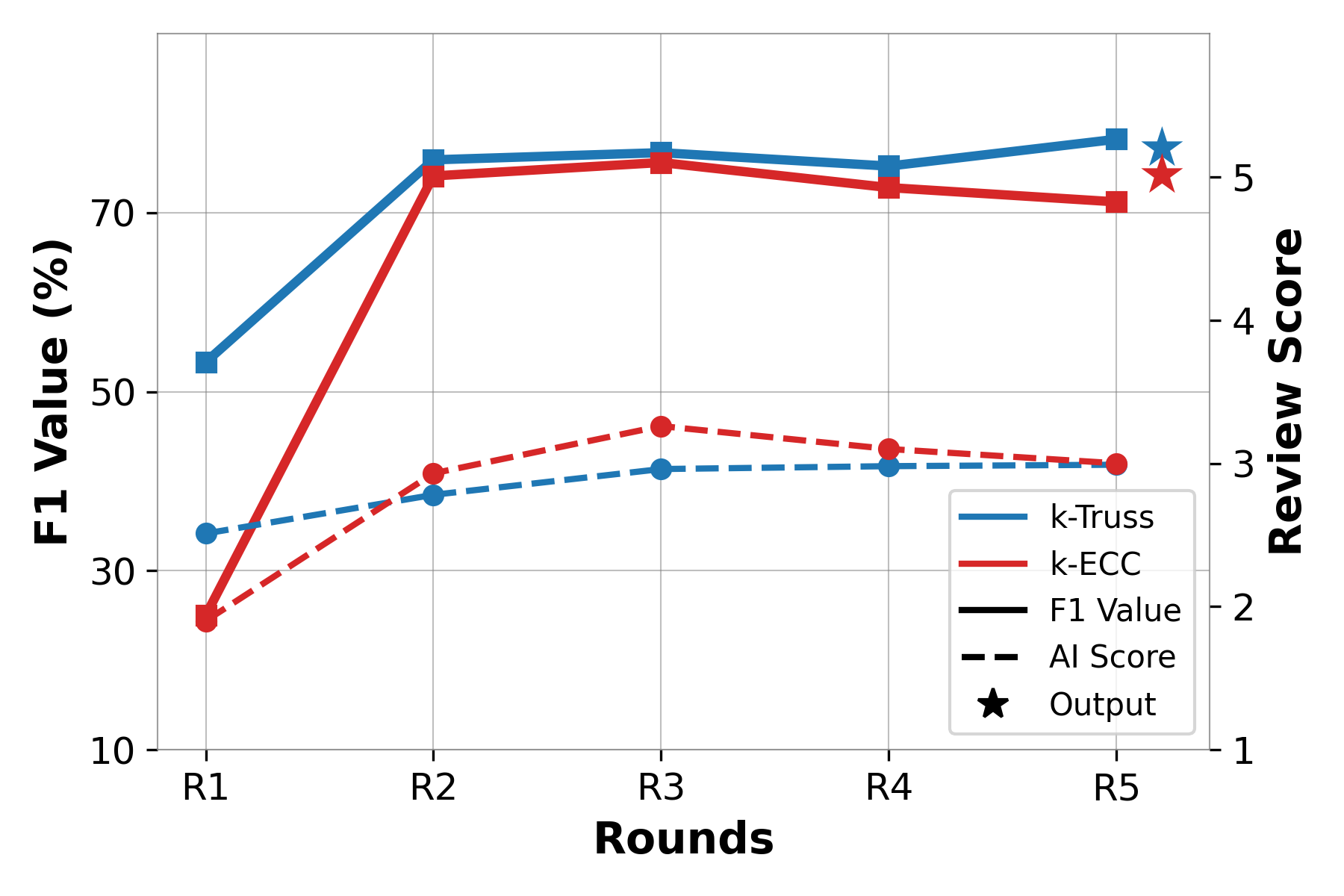}
        \caption{$k$-Truss and $k$-ECC}
        \label{fig:zero_truss_ecc}
    \end{subfigure}
    
    
    
    \caption{Bias between output community's F1 value and the review score given by the Validator in each round over four tasks. (a) $k$-Core and $k$-Clique. (b) $k$-Truss and $k$-ECC. The F1 value of the output community is marked by star.}
    \label{fig:bias}
\end{figure}

\decider module ensures the overall performance of \csagent by determining the output community. In order to examine whether \decider can make correct choices among multiple communities, we compared the F1 value of the final output community (marked by star) with the communities obtained in each round of dialogue, as depicted in Figure \ref{fig:bias}. The results demonstrate that the \decider is capable of making appropriate choices in most cases, leveraging community characteristics and evaluation feedback from the \validator to guide its decisions. Notably, in $k$-Clique task, the selected community achieved an F1 value even surpass the best F1 value obtained in any single round. In other tasks, F1 value of the output communities are also nearly equivalent to the highest F1 scores, not affected much by the decline in community quality as the number of rounds increases. These emphasize the capability of \decider module to leverage both community features and \validator feedback to make informed decisions, thereby ensuring that the final output closely approximates or even exceeds the best results obtained during the iterative dialogue process.

\subsubsection{Advantages of Multi-round Refinement} 
The principal idea of \csagent is to obtain candidate communities through multi-rounds of dialogue between two agents, and then select the best one as the output through the \decider module. Existing researches have also improved the performance of \llms in reasoning tasks by generating multiple results, among which self-consistency (SC) method is the most representative one \cite{wang2022self}. The SC method enhances robustness of Zero-shot prompting method by generating multiple outputs. To validate the advantages of multi-round refinement component in \csagent, we conduct a comparative analysis with the SC method. Assuming \csagent involves $k$ rounds of dialogue, we employed the SC method to generate $k$ candidate communities. For each vertex $n_i$, we counted its occurrence $c_i$ across the $k$ communities. A vertex was included in the final output community if it satisfied the majority voting criterion $ c_i \times 2 \geq k $.

Results in Table \ref{tab:sc} reveal that the SC method does not meet expectations in SC tasks. Zero-shot based \csagent outperforms SC methods on all tasks. The SC method performs close to \csagent only in some easy tasks (e.g., 3.7\% lower in $k$-Clique on LFR dataset), and is far inferior to \csagent in other aspects (e.g., 67.4\% lower in $k$-ECC on PSG dataset). This highlights the critical role of multi-round refinement in \csagent, which cannot be substituted by merely generating multiple paths or answers. The iterative dialogue process ensures the unique advantage of \csagent in achieving more reliable and accurate results.

\subsubsection{Better with more rounds?} 
\begin{figure}
    \centering
    \captionsetup[subfigure]{font=footnotesize}
    \begin{subfigure}[b]{0.23\textwidth}
        \centering
        \includegraphics[width=\linewidth]{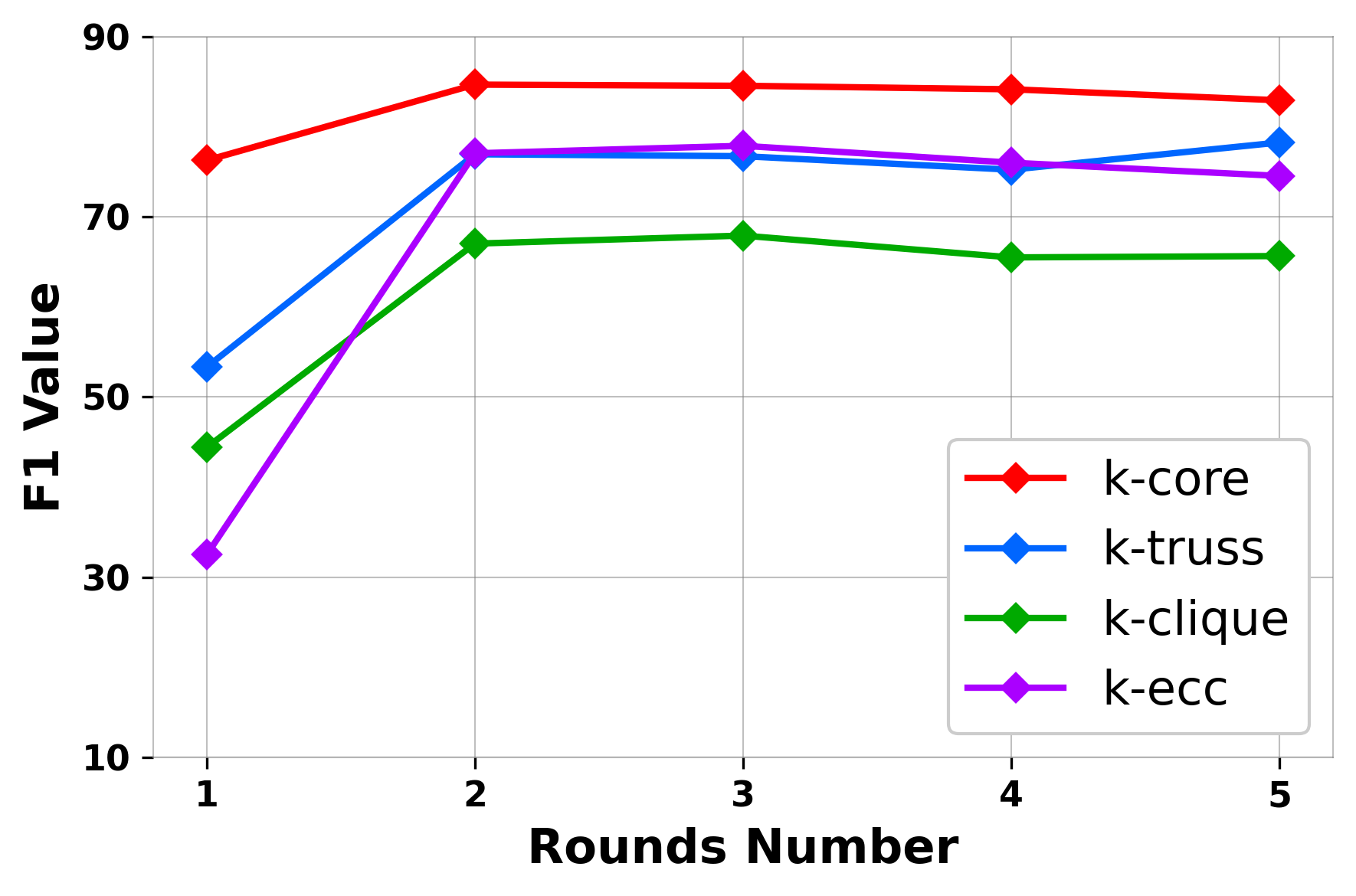}
        \caption{Trends on Random Dataset}
        \label{fig:round_left}
    \end{subfigure}
    \hspace{0.005\textwidth}
    \begin{subfigure}[b]{0.23\textwidth}
        \centering
        \includegraphics[width=\linewidth]{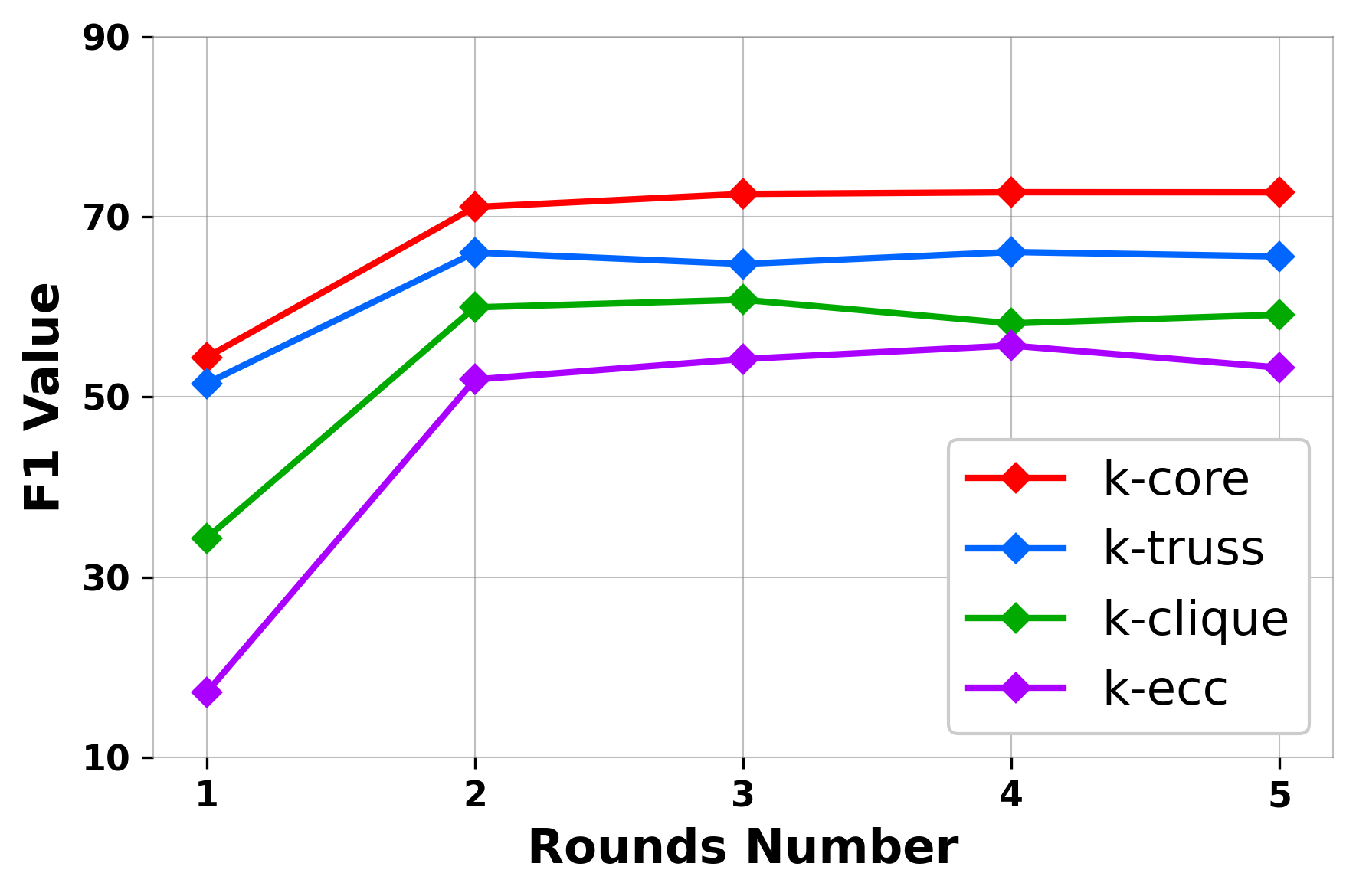}
        \caption{Trends on LFR Dataset} 
        \label{fig:round_right}
    \end{subfigure}
    \caption{Performance trends of different algorithms in multi-round dialogues.}
    \label{fig:multi_round}
\end{figure}

To examine the impact of iterative refinement depth on model performance, we conduct controlled experiments with the dialogue rounds parameter $r=5$. As illustrated in Figure \ref{fig:multi_round}, we quantitatively compared the F1 scores of identified communities across different rounds. Our analysis reveals that the performance gains from additional dialogue rounds exhibit diminishing returns, with marginal improvements observed beyond $r=3$. More notably, certain tasks (particularly $k$-ECC and $k$-Clique across both datasets) demonstrate performance degradation with increased rounds, suggesting limitations in current \llms' self-reflection capabilities \cite{huang2023large}. Therefore, increasing the number of dialogue rounds does not necessarily improve the overall performance of \csagent. Additionally, from a computational efficiency perspective, each additional round linearly increases token consumption by approximately 25-30\%, while providing negligible quality improvements. Based on this systematic analysis, we ultimately selected $r=3$ as the optimal configuration when conducting experiments, achieving a balanced trade-off between performance and computational overhead.

\section{RELATED WORK}

\subsection{LLMs on Graph Data Mining}
A series of benchmarks have extensively evaluated \llms' capabilities in graph reasoning and structure understanding \cite{guo2023gpt4graph, he2024g, chen2024exploring, tang2024graphgpt, li2024can, wang2024can}. Among these, Taylor et al. \cite{taylor2024large} propose MAGMA to assess \llms on classical graph algorithms (e.g., BFS and Dijkstra). Their findings reveal significant challenges \llms encounter in executing these algorithms step-by-step. Similarly, Dai et al. \cite{dai2024large} demonstrated \llms' preliminary graph pattern recognition abilities, with improved performance when input aligns with pretraining data. Further advancing this field, ComGPT \cite{ni2024comgpt} is a GPT-guided framework for local community detection,
which outperforms traditional methods by addressing seed dependency and diffusion issues. Collectively, these works emphasize \llms' potential in graph tasks while underscoring the need for deeper exploration of complex graph mining and optimized model development.

\subsection{LLMs on Self-correction}
Recent studies reveal an intriguing ability of \llms: self-correction, where they can iteratively refine their initial responses based on feedback. REFINER \cite{paul2023refiner} and SELF-REFINE \cite{madaan2023self} are the most influential works that employ iterative feedback loops to enhance \llms' outputs in dialogue generation and mathematical reasoning. Evoke \cite{hu2023evoke} and Multi-Agent Debate (MAD) \cite{liang2023encouraging} utilize multi-agent frameworks to refine outputs. They achieve excellent results in tasks such as arithmetic reasoning and machine translation. Moreover, Reflexion \cite{shinn2023reflexion} enables \llms to reflect on feedback verbally and store insights in episodic memory, improving performance across tasks like coding and language reasoning. Self-Contrast \cite{selfself} mitigates overconfidence and randomness by exploring diverse perspectives and summarizing discrepancies, which improves performance on reasoning and translation tasks. In summary, these findings highlight that \llms possess measurable self-correction capabilities, particularly in NLP tasks.

\section{CONCLUSION AND FUTURE WORK}
In this paper, we explored the potential of \llms in addressing the community search problem, a fundamental task in graph analysis. We introduce \graphcs, a comprehensive benchmark that evaluates \llms on diverse graph structures. The evaluation reveals their unsatisfactory performance and the presence of output bias. To overcome these limitations, we propose a dual-agent collaborative framework \csagent, which leverages the complementary roles of \solver and \validator \llms to iteratively refine community search results. \decider module further ensures the selection of optimal communities. Experimental results demonstrate that \csagent significantly outperforms baseline methods in both quality and stability. Our work provides a robust and adaptive \llm-based solution for CS tasks across various domains. 

In future work, we will first expand our \graphcs benchmark, process complex real-world data sets, enable it to be applied to \llm-based CS tasks, and introduce more complex community search tasks (such as not giving a specific metric and letting \llms explore the potential community structure themselves). Secondly, we will continue to optimize the \csagent framework, further expand the structure of the agent by introducing external corpus, thus stimulating the self-refection potential of \llms.

\begin{acks}
Long Yuan is supported by NSFC62472225. The authors acknowledge the use of DeepSeek for improving readability and grammar checking. The authors independently verified all outputs and assume full responsibility for the manuscript’s content. 
\end{acks}

\printbibliography

\appendix

\end{document}